\documentclass[11pt]{article}
\usepackage[top=1in, bottom=1.2in, left=1.1in, right=1.1in]{geometry}
\usepackage{setspace}
\usepackage{amsmath,amsfonts,amssymb}
\usepackage{amsthm}
\usepackage{color}
\usepackage{bm}
\usepackage{natbib}
\usepackage{appendix}
\usepackage{float}
\usepackage{graphicx}
\usepackage{enumitem}
\usepackage{multirow}
\usepackage{authblk}
\usepackage[breaklinks=true]{hyperref}
\usepackage{booktabs}
\usepackage[flushleft]{threeparttable}
\usepackage{algorithm}
\usepackage{algorithmic}
\usepackage{caption}
\usepackage{threeparttable}
\usepackage{subcaption}
\usepackage{amssymb}
\usepackage{diagbox}
\usepackage{graphicx}
\usepackage{latexsym}
\usepackage{amsfonts}
\usepackage{color}
\usepackage{amsmath}
\usepackage{multicol}
\usepackage{float}
\usepackage{hyperref}
\usepackage{graphicx}
\usepackage{ulem}
\usepackage{threeparttable}
\usepackage{multirow}
\usepackage{enumitem}
\usepackage{epstopdf}
\usepackage{bm}
\usepackage{amsmath, amssymb}
\usepackage{algorithm}
\usepackage{algorithmic}
\usepackage{makecell}
\usepackage{rotating} 
\usepackage{float}
\usepackage{subcaption}
\usepackage{tikz}

\csname @openrightfalse\endcsname

\setlist{nolistsep}
\makeatletter
\newcommand*{\rom}
[1]{\expandafter\@slowromancap\romannumeral #1@}
\makeatother

\newtheorem{theorem}{Theorem}
\newtheorem{proposition}{Proposition}

\newtheorem{lemma}{Lemma}

\setenumerate{noitemsep}

\hypersetup{
	colorlinks   = true,
	citecolor    =  blue
}

\usepackage{titlesec}

\newtheorem{corollary}{Corollary}

\title{Efficient Covariance Estimation for Sparsified Functional Data}

\author[1]{Sijie Zheng\thanks{Work was done during an internship at the Pattern Recognition Center, WeChat AI, Tencent Inc., China.}}
\author[2]{Fandong Meng}
\author[2]{Jie Zhou}

\affil[1]{Center for Statistical Science and Department of Industrial Engineering, Tsinghua University}
\affil[2]{Pattern Recognition Center, WeChat AI, Tencent Inc., China}

\affil[ ]{\texttt{zhengsj0212@gmail.com}}

\date{}
\begin{document}
\maketitle
\begin{abstract}
Motivated by recent work involving the analysis of leveraging spatial correlations in sparsified mean estimation, we present a novel procedure for constructing covariance estimator.
The proposed Random-knots (Random-knots-Spatial) and B-spline (Bspline-Spatial) estimators of the
covariance function are computationally
efficient. Asymptotic pointwise of the covariance
are obtained for sparsified individual trajectories under some regularity conditions. Our proposed nonparametric method well perform the functional principal components analysis for the case of sparsified data, 
where the number of repeated measurements available per subject is small. In contrast, classical functional data analysis requires a large number of regularly spaced measurements per subject. Model selection techniques, such as the Akaike information criterion, are used to
choose the model dimension corresponding to the number of eigenfunctions in the model. Theoretical results are illustrated with Monte Carlo simulation experiments. Finally, we cluster multi-domain data by replacing the covariance function with our proposed covariance estimator during PCA.
\end{abstract}

\section{Introduction}
The goal of this paper is to estimate the covariance function of a set of vectors collected from a distributed system of nodes. Dimension reduction has received increasing attention so as to avoid expensive and slow computation. \cite{Stich18} has investigated the converges rate of Stochastic Gradient Descent (SGD) after applying sparsification or compression. \cite{Jhun21} has focused on the mean function of a sparsified vector containing only a subset of the elements of the original vector.
In many practical applications, the vectors sent by the nodes are correlated across different nodes and over consecutive rounds of iterative algorithms.

Data takes the form $\left\{\left(x_{i j}, j / d\right), 1 \leq i \leq n, 1 \leq j \leq d\right\}$
in which $x_{i}(\cdot)$ is a latent smooth trajectory for subject $i$,
\begin{align}
\label{model}
x_{i}(\cdot)=m(\cdot)+Z_{i}(\cdot),
\end{align}
where the deterministic function $m(\cdot)$ denotes the common population mean, the random $Z_{i}(\cdot)$ subject-specific small variation with $\mathbb{E} Z_{i}(\cdot) \equiv 0$.

The trajectories $x_{i}(\cdot)$ are identically distributed realizations of $\{x(t), t \in[0,1]\}$, a continuous stochastic process defined on $[0,1]$, with $\mathbb{E} \sup _{t \in[0,1]}|x(t)|^{2}<+\infty$. $x(\cdot)$ can be decomposed as $x(\cdot)=m(\cdot)+Z(\cdot)$, where $Z\left(\cdot\right)$ satisfies $\mathbb{E} Z(t)=0$. The covariance function of $x(\cdot)$ is denoted by $G\left(t, t^{\prime}\right)=\operatorname{Cov}\left\{x(t), x\left(t^{\prime}\right)\right\}=$ $\operatorname{Cov}\left\{Z(t), Z\left(t^{\prime}\right)\right\}, t, t^{\prime} \in[0,1]$. Functions $m\left(\cdot\right)$ and $Z_{i}\left(\cdot\right)$ are viewed as smooth functions of time $t=\frac{j}{d}$ for $j=1,\ldots,d$, which is rescaled to domain $[0, 1]$.

Let sequences $\left\{\lambda_{k}\right\}_{k=1}^{\infty}$ and $\left\{\psi_{k}\right\}_{k=1}^{\infty}$ be the eigenvalues and eigenfunctions of $G\left(t, t^{\prime}\right)$, respectively, in which $\lambda_{1} \geq \lambda_{2} \geq \cdots \geq 0, \sum_{k=1}^{\infty} \lambda_{k}<\infty$, $\left\{\psi_{k}\right\}_{k=1}^{\infty}$ form an orthonormal basis of $L^{2}[0,1]$. Mercer Lemma entails that the $\psi_{k}$ 's are continuous and continuous covariance function $G\left(t, t^{\prime}\right)=\sum_{k=1}^{\infty} \lambda_{k} \psi_{k}(t) \psi_{k}\left(t^{\prime}\right), \int G\left(t, t^{\prime}\right) \psi_{k}\left(t^{\prime}\right) d t^{\prime}=\lambda_{k} \psi_{k}(t), t, t^{\prime} \in[0,1]$. The standard process $x(\cdot)$ allows Karhunen-Loève $L^{2}$ representation $x(\cdot)=m(\cdot)+$ $\sum_{k=1}^{\infty} \xi_{k} \phi_{k}(\cdot)$, in which the random coefficients, $\xi_{k}$, called functional principal component (FPC) scores, are uncorrelated with mean 0 and variance 1. The rescaled eigenfunctions, $\phi_{k}$, called FPC, satisfy $\phi_{k}=\sqrt{\lambda_{k}} \psi_{k}$ and $\int\{x(t)-m(t)\} \phi_{k}(t) d t=\lambda_{k} \xi_{k}$, for $k \geq 1$.  Although the
sequences $\left\{\lambda_{k}\right\}_{k=1}^{\infty}$, $\left\{\phi_{k}\right\}_{k=1}^{\infty}$ and $\left\{\xi_{ik}\right\}_{i=1, k=1}^{n,\infty}$ exist mathematically, they are either unknown or unobservable.
\subsection{Main contribution}
The sparsity of the vector $x_{i}=\left(x_{i1}, \ldots, x_{id}\right)$ corresponding to the $i$-th node is determined by $J_{s}/d$, and the value of $J_{s}$ is determined by the dimension and fluctuation of the original vector. Akaike information criterion (AIC) is applied and simulation results reveal the superiority of this data-driven method comparing to other sparse method with fixed number of nodes. 

\cite{Jhun21} have proposed a ``Random-knots'' sparse method and computed the accuracy of the sample mean function $\hat{m}$ obtained by the sparsified vectors $\left\{h_{i}\right\}$ compared with the mean estimator $\bar{m}$ using the original data $\left\{x_{i}\right\}$. On the basis of their research, we further propose the estimation covariance function $\hat{G}\left(\cdot, \cdot\right)$ under the ``Random-knots'' sparse method, and prove that $\hat{G}\left(\cdot, \cdot\right)$ has good properties and can effectively approximates the sample covariance function $\bar{G}\left(\cdot, \cdot\right)$ computed from the original data. With carefully chosen function $T(\cdot)$ to describe correlation between nodes, our Random-knots-Spatial estimator has the potential to be better than Random-knots estimator, i.e. smaller mean squared error (MSE). We figured out the optimal $T^{*}(\cdot)$ for particular data set $\left\{x_{ij}\right\}_{i=1, j=1}^{n, d}$. 

For $J_{s}$ nodes of equispaced distribution, we propose a B-spline interpolation method to describe the correlation between the dimensions of vectors, which can also be viewed as temporal correlation. Compared with the sparsified vector obtained by method ``Random-knots'', the sparsified vector $\left\{h_{i}\right\}$ obtained by this two-step estimation method ``B-spline'' can effectively approximate original trajectory $\left\{x_{i}\right\}$. The B-spline (Bspline-Spatial) covariance estimator has globally consistent convergence rate, enjoying better theoretical properties than the estimator without interpolation.

The paper is organized as follows. Section \ref{SEC:result} introduces four kinds of two-step B-spline covariance estimtor and proves that the
proposed estimator is asymptotically equivalent to sample estimator without sparsification. Our covariance estimators are guaranteed to be the positive semi-definite and estimator error can be drastically reduced when there are spatial and temporal correlations. Section \ref{SEC:fpc} deduces the convergence rate of FPC and FPC score based on the convergence rate of covariance function. More
detailed implementation procedures are presented in Section \ref{SEC:imp}. We present the
simulation studies in Section \ref{SEC:sim} and applications to machine translation in Section \ref{SEC:app}. All technical proofs are involved in the Appendix.

\subsection{Related work}
Considerable efforts have been made to analyze first-order structure of function-valued random elements, i.e., the functional mean $m\left(\cdot\right)$. Estimation of mean function has been investigated in \cite{Jhun21}, \cite{Garg14}, \cite{Sur17}, \cite{May21} and \cite{Brown21}, \cite{zheng2025inference}. Works \cite{Cao12} and \cite{Huang22} consider empirical
mean estimation using B-spline estimation. The second-order structure of random functions in the covariance
function $G\left(\cdot, \cdot\right)$, the next object of interests. \cite{Cao16}, \cite{Zhong22} and \cite{zheng2025inference2} have proposed tensor product B-spline covariance estimator. To the best of our knowledge, spatial correlation
across nodes has not yet been considered in the context of sparsified covariance estimation. The research on Sparsification has received more and more attention recently, for instance \cite{Ali18}, \cite{Stich18} and \cite{Sahu21}. Sparsification methods mainly focus on sending only a subset of elements of the
vectors, yet no existing method combine sparsity method with B-spline fittin. There have been striking
improvement over sparse PCA. \cite{Bert132} and \cite{Choo21} have analyzed the complexity of sparse PCA; \cite{Bert13} and \cite{Des14} have obtained sparse principle components for particular data models. We study the convergence rate of eigenvalues and eigenfunctions of our covariance estimators. Since our four estimation methods is original, the corresponding study of PCA is proposed for the first time. 

\section{Main results}
\label{SEC:result}
We consider $n$ geographically distributed nodes, each node generates a $d$-dimensional vector $x_{i}=\left[x_{i 1}, \ldots, x_{i d}\right]^{\top}$ for $i=1, \ldots, n$. The mean function $m \left(\cdot\right)$ could be estimated by simply average samples of all nodes, for $t, t^{\prime}\in [0, 1]$,
\begin{align}
\label{m}
\bar{m}(t)=\frac{1}{n} \sum_{i=1}^{n} x_{i}(t)
\end{align}
and the averaged covariance estimator
\begin{align}
\label{cov2}
\bar{G}\left(t, t^{\prime}\right)=\frac{1}{n}\sum_{i=1}^{n}\left(x_{i}\left(t\right)-\bar{m}\left(t\right)\right)\left(x_{i}\left(t^{\prime}\right)-\bar{m}\left(t^{\prime}\right)\right)
\end{align}

For a non-negative integer $q$ and a real number $\mu \in(0,1]$, write $\mathcal{H}^{(q, \mu)}[0,1]$ as the space of $\mu$-Hölder continuous functions, i.e.,
$$
\mathcal{H}^{(q, \mu)}[0,1]=\left\{\varphi:[0,1] \rightarrow \mathbb{R} \mid\|\varphi\|_{q, \mu}=\sup _{x, y \in[0,1], x \neq y} \frac{\left|\varphi^{(q)}(x)-\varphi^{(q)}(y)\right|}{|x-y|^{\mu}}<+\infty\right\}
$$
We next introduce some technical assumptions.

\textbf{Assumption 1:} There exists an integer $q>0$ and a constant $\mu \in(0,1]$, such that the regression function $m(\cdot) \in \mathcal{H}^{(q, \mu)}[0,1]$. In the following, we denote $p^{*}=q+\mu$ for simplicity.

\textbf{Assumption 2:} The covariance function satisfies $\sup _{\left(t, t^{\prime}\right) \in[0,1]^{2}} G\left(t, t^{\prime}\right)<C$, for some positive constant $C$ and $\min _{t \in[0,1]} G\left(t, t^{\prime}\right)>0$.

\textbf{Assumption 3:} There exists a constant $\theta>0$, such that as $d \rightarrow \infty, n=n(d) \rightarrow \infty, n=\mathcal{O}\left(d^{\theta}\right)$.

\textbf{Assumption 4:} The rescaled FPCs $\phi_{k}(\cdot) \in \mathcal{H}^{(q, \mu)}[0,1]$ with $\sum_{k=1}^{\infty}\left\|\phi_{k}\right\|_{q, \mu}<+\infty$, $\sum_{k=1}^{\infty}\left\|\phi_{k}\right\|_{\infty}<+\infty$; for increasing positive integers $\left\{k_{n}\right\}_{n=1}^{\infty}$, as $n \rightarrow \infty, \sum_{k_{n}+1}^{\infty}\left\|\phi_{k}\right\|_{\infty}=\mathcal{O}\left(n^{-1 / 2}\right)$ and $k_{n}=\mathcal{O}\left(n^{\omega}\right)$ for some $\omega>0$.

\textbf{Assumption 5:} The FPC scores $\left\{\xi_{i k}\right\}_{i \geq 1, k \geq 1}$ are independent over $k \geq 1$ and i.i.d over $i \geq 1$. The number of distinct distributions for all FPC scores $\left\{\xi_{i k}\right\}_{i \geq 1, k \geq 1}$ is finite, and $\max _{1 \leq k<\infty} \mathbb{E} \xi_{1 k}^{r_{0}}<\infty$ for $r_{0}>4$.

Assumptions 1--5 are standard requirements for obtaining the mean and covariance estimators in literature. Assumption 1 guarantee the orders of the bias
terms of the spline smoothers for $m\left(\cdot\right)$. Assumption 2 ensures the covariance $G\left(\cdot, \cdot\right)$ is a uniformly bounded function. Assumption 3 implies the number
of points on each curve $d$ diverges to infinity as $n\rightarrow \infty$, which is a well-developed
asymptotic scenario for dense functional data. It is important to notice that this assumption is practically relevant since curves or images
measured using new technology are usually of much higher resolution than the
previous generation. This directly leads to the doubling of the amount of data recorded in each node, which is also the motivation of this paper to propose sparsification before feature extraction, modeling, or other steps. Assumption 4
concerns the bounded smoothness of FPC and Assumption 5 ensures bounded FPC scores, for bounding the bias
terms in the spline covariance estimator.

The smoothness of our estimator is controlled by the number
of knots, which increases to infinity as specified in Assumption 6. This increasing knots
asymptotic framework guarantees the richness of the basis. We design two sparsification methods: Random-Sparsification and Fix-Sparsification, where Random-Sparsification randomly selects $J_{s}$ component in the $d$-dimensional vector of each node, and Fix-Sparsification selects $J_{s}$ component at fixed position. The proportion $\frac{J_{s}}{d}$ depicts the difference of data volume before and after sparsification, reflecting the degree of sparsification. The value of $J_{s}$ is completely determined by the original data.

\textbf{Assumption 6:} The number of interior knots $J_{s} \asymp d^{\gamma} C_{d}$ for some $\tau>0$ with $C_{d}+C_{d}^{-1}=\mathcal{O}\left(\log ^{\tau} d\right)$ as $d \rightarrow \infty$, and for
Random-Sparsification: $\gamma\geq 1-\frac{\theta}{2}$; for Fix-Sparsification: $\gamma>\frac{\theta}{2p^{*}}+\frac{2\theta}{r_{0}p^{*}}$.

These assumptions are mild conditions that can be satisfied in many practical situations. One simple
and reasonable setup for the above parameters $q$, $\mu$, $\theta$, $p$, $\gamma$ can be as follows: $q + \mu = p^{*} = 4$, $\theta = 1$,
$p = 4$, $\gamma = 5/8$, $C_{d} \asymp \log\log d$. These constants are used as defaults to construct candidate pool for the number of knots $J_{s}$ in Section \ref{AIC} for two sparsification schemes.

\subsection{Random-Sparsification}
In the case of randomly selected points for sparsification at each node, we propose two kinds of estimation of covariance function: Random-knots and Random-knots-Spatial estimator, where the samples of constructing Random-knots estimator at different nodes are assumed to be generated independently, and Random-knots-Spatial estimator further takes the correlation between nodes into account on the basis of Random-knots estimator.

{\textbf{Random-knots estimator}} For node $i$, $h_{i}=\left[h_{i 1}, \ldots, h_{i d}\right]^{\top}$ is a sparsified version of the corresponding data vector $x_{i}=\left[x_{i 1}, \ldots, x_{i d}\right]^{\top}$. The estimator generated from $h_{1}, \ldots, h_{n}$ is called sparsified estimator. If $\left\{h_{i}\right\}_{i=1}^{n}$ are sparsified vectors randomly containing $J_{s}$ elements of the original vector, i.e. $\mathbb{P}\left(h_{ij}=0\right)=1-\frac{J_{s}}{d}$, $\mathbb{P}\left(h_{ij}=x_{ij}\right)=\frac{J_{s}}{d}$, we obtain the Random-knots mean and covariance estimator by replacing $x_{i}$ in (\ref{m}), (\ref{cov2}) by $h_{i}$,
\begin{align}
&\hat{m}=\frac{1}{n} \frac{d}{J_{s}} \sum_{i=1}^{n}h_{i}\label{Random-k mean}\\
&\hat{G}=\frac{1}{n}\left(\frac{d}{J_{s}}\right)^{2}\sum_{i=1}^{n}\left(h_{i}-\bar{h}\right)\left(h_{i}-\bar{h}\right)^{\top}\label{Random-k cov}
\end{align}
The following theorem states the mean squared error (MSE) of the proposed covariance estimator $\mathbb{E}\|\hat{G}-\bar{G}\|^{2}$ tends to zero as the number of nodes goes to infinity. The proof is
analogous to Lemma $1$ in \cite{Jhun21} where MSEs of mean $\hat{m}$ and covariance $\hat{G}$ are defined as
\begin{align*}
\operatorname{MSE}_{m}=\frac{1}{d} \sum_{j=1}^{d}\left(\bar{m}_{j}-\hat{m}_{j}\right)^{2},\quad\operatorname{MSE}_{G}=\frac{1}{d^{2}} \sum_{j, j^{\prime}=1}^{d}\left(\bar{G}_{j j^{\prime}}-\hat{G}_{j j^{\prime}}\right)^{2}
\end{align*}

\begin{figure}[!h]
    \centering
    \includegraphics[width=\linewidth]{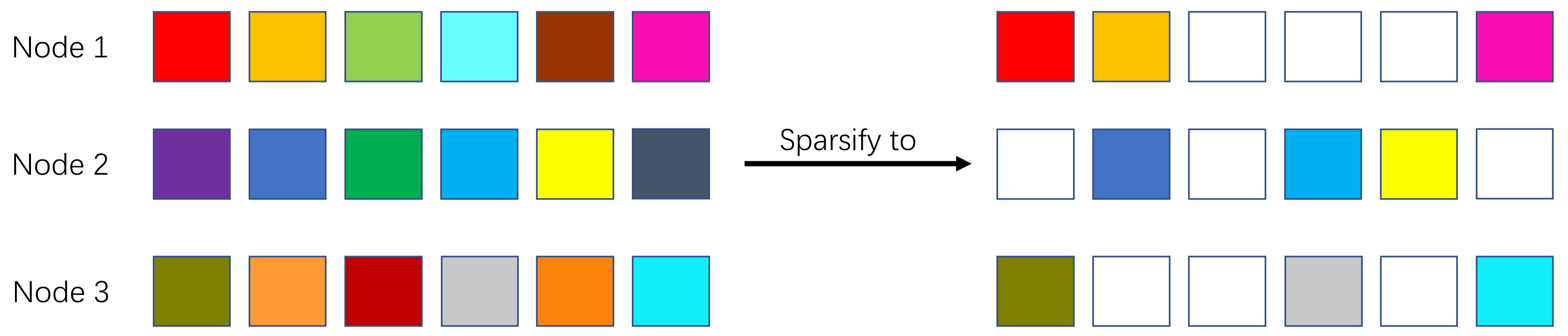}
    \caption{The Random-knots method to estimate mean and covariance function with $n=3$ number of nodes, left: origin vectors $\left\{x_{i}\right\}_{i=1}^{3}$ with dimension $d=6$, right: sparsified vectors $\left\{h_{i}\right\}_{i=1}^{3}$ and the number of randomly chosen non-zero elements for each vector is $J_{s}=3$.}
\end{figure}
\begin{theorem}
\label{Random-k}
(Random-knots Estimation Error). Under Assumptions 1--5, MSE of estimate $\hat{G}$ produced by the Random-knots sparsification scheme described above is given by
$$
\mathbb{E}\|\hat{G}-\bar{G}\|^{2}=\frac{1}{n^{2}}\left(\left(\frac{d}{J_{s}}\right)^{2}-1\right) R_{1}
$$
where $R_{1}=\sum_{i=1}^{n}\left\|x_{i}-\bar{m}\right\|^{4}$, the sum of the squared magnitudes of the data vectors.

Assumption 6 further guarantees that
$
\left\|\hat{G}-\bar{G}\right\|_{2}=\mathcal{O}_{p}\left(n^{-1/2}\right).
$
\end{theorem}

\textbf{Random-knots-Spatial estimators} To further increase the accuracy of Random-knots estimators, \cite{Jhun21} has introduced $M_{j}$ to describe the correlation between nodes. $M_{j}$ is the number of nodes that send their $j$-th coordinate. It is obvious that $M_{j}$ is a binomial random variable that takes values in the range $\left\{0, 1, \ldots , n\right\}$
with $\mathbb{P}\left(M_{j} = m\right) = \binom{n}{m}p^{m}\left(1-p\right)^{n-m}$ with $p=\frac{J_{s}}{d}$. If $M_{j}=0$, it means that none of the $n$ nodes has drawn the $j$-th element, and the information about position $j$ is completely missing.

If the vectors of $n$ nodes are highly correlated, we can still obtain an accurate estimator even if few points are selected at position $j$. Consider a special case where each node has the same vector, i.e., $x_{1} = x_{2} = \ldots = x_{n}$. The $j$-th coordinate of mean can be exactly estimated as $\hat{m}_{j}=\frac{1}{M_{j}}\sum_{i=1}^{n}h_{ij}$ whenever $M_{j}>0$. And the exact variance function at position $j$ is $\hat{G}_{jj^{\prime}}=\frac{1}{M_{j}}\sum_{i=1}^{n}\left(h_{ij}-\bar{h}_{j}\right)^{2}$, the covariance function is $\hat{G}_{jj^{\prime}}=\frac{1}{M_{j}}\sum_{i=1}^{n}\left(h_{ij}-\bar{h}_{j}\right)\left(h_{ij^{\prime}}-\bar{h}_{j^{\prime}}\right)$, $j^{\prime}\neq j$. Simple mathematical derivation implies that $\bar{h}_{j}=\hat{m}_{j}$ under this situation. Hence, the scaling parameter $\frac{J_{s}}{d}$ in (\ref{Random-k}) is not necessary. 

However, if the correlation between nodes is low, too small $M_{j}$ value may lead to a large MSE of estimator. For instance, the $d$-dimension vector corresponding to the other $n-1$ nodes is a sine distribution of dimension $x_{ij}=\sin\left(2\pi \frac{j}{d}\right)$, $1\leq i\leq n-1$, $1\leq j\leq d$ and only one node is a cosine distribution $x_{nj}=\cos\left(2\pi \frac{j}{d}\right)$ or the vector has a jump in the $j$-th component $x_{nj}=\sin\left(2\pi \frac{j}{d}\right)+\delta$, $x_{nj^{\prime}}=\sin\left(2\pi \frac{j^{\prime}}{d}\right)$, $\delta>0$, $j^{\prime}\neq j$, which can be regarded as an outlier. In special case that only the point of cosine distribution is selected for position $j$, the estimation of this position is bound to have a large deviation. Therefore, the correlation between nodes is an important indicator to measure the accuracy of estimator.

We propose the Random-knots-Spatial estimator
wherein the fix scaling parameter $\frac{J_{s}}{d}$ is replaced by a function of $M_{j}$ such that the level of spatial correlation between the
vectors is taken into account. In particular, the mean estimator for $j$-th element is
\begin{align}
\label{Random-k spatial mean}
\hat{m}_{j}=\frac{1}{n} \frac{\bar{\beta}}{T\left(M_{j}\right)} \sum_{i=1}^{n} h_{i j},
\end{align}
And covariance function at position $\left(j, j^{\prime}\right)$ is
\begin{align}
\label{Random-k spatial cov}
\hat{G}_{jj^{\prime}}=&\frac{1}{n}\left(\frac{d}{J_{s}}\right)^{2}\frac{\left(\mathbb{E}_{M_{j} \mid M_{j} \geq 1}\left(\frac{1}{T\left(M_{j}\right)}\right)\mathbb{E}_{M_{j^{\prime}} \mid M_{j^{\prime}} \geq 1}\left(\frac{1}{T\left(M_{j^{\prime}}\right)}\right)\right)^{-1}}{T\left(M_{j}\right)T\left(M_{j^{\prime}}\right)}\sum_{i=1}^{n}\left(h_{ij}-\bar{h}_{j}\right)\left(h_{ij^{\prime}}-\bar{h}_{j^{\prime}}\right)\notag\\
=&\frac{1}{n} \frac{\bar{\beta}^{2}}{T\left(M_{j}\right)T\left(M_{j^{\prime}}\right)}\sum_{i=1}^{n}\left(h_{ij}-\bar{h}_{j}\right)\left(h_{ij^{\prime}}-\bar{h}_{j^{\prime}}\right)
\end{align}
where function $T\left(M_{j}\right)$ changes the scaling of $h_{i j}$ depending on $M_{j}$, the number of nodes that send the $j$-th coordinate. And $\bar{\beta}$ is 
\begin{align}
\label{def:betabar}
\bar{\beta} = &\left(\frac{J_{s}}{d} \mathbb{E}_{M_{j} \mid M_{j} \geq 1}\left(\frac{1}{T\left(M_{j}\right)}\right)\right)^{-1}\notag\\
=&\left(\sum_{r=1}^{n} \frac{J_{s}}{d T(r)}\binom{n-1}{r-1}\left(\frac{J_{s}}{d}\right)^{r-1}\left(1-\frac{J_{s}}{d}\right)^{n-r}\right)^{-1}
\end{align}
\cite{Jhun21} has proved that the Random-knots-Spatial mean estimator in (\ref{Random-k spatial mean}) is unbiased. On their basis, the following proposition proves the unbiasedness of covariance estimator.

\begin{proposition}
\label{THM:Random-k-spatial 1}
(Random-knots-Spatial estimator Unbiasedness). Under Assumptions 1--6, we have
$$\mathbb{E}\hat{G}=\bar{G}.$$
\end{proposition}
This proposition is equivalent to prove
$\mathbb{E}\hat{G}_{jj^{\prime}}=\frac{1}{n} \sum_{i=1}^{n} G_{i j j^{\prime}}$. One trick is to introduce the random variable $\xi_{ij}$ to aid in the computation of conditional expectations $\mathbb{E}_{M_{j} \mid M_{j}\geq 1}\left(\cdot\right)$. Indicator $\xi_{i j}$ is $1$ or $0$, depending on whether $h_{i j}=x_{i j}$ or not, leading to $\mathbb{E}_{M_{j} \mid M_{j}\geq 1, M_{j^{\prime}}\geq 1}\left(\cdot\right)=\mathbb{E}_{M_{j} \mid \xi_{ij}=1, \xi_{ij^{\prime}}=1}\left(\cdot\right)$, $\mathbb{E}_{M_{j} \mid M_{j}= 0, M_{j^{\prime}}\geq 1}\left(\cdot\right)=\mathbb{E}_{M_{j} \mid \xi_{ij}=0, \xi_{ij^{\prime}}=1}\left(\cdot\right)$, $\mathbb{E}_{M_{j} \mid M_{j}=0, M_{j^{\prime}}=0}\left(\cdot\right)=\mathbb{E}_{M_{j} \mid \xi_{ij}=0, \xi_{ij^{\prime}}=0}\left(\cdot\right)$. 
Event $\left\{\xi_{i j}=1, \xi_{i j^{\prime}}=1\right\}$ happens with probability $\left(\frac{J_{s}}{d}\right)^{2}$, event $\left\{\xi_{i j}=0, \xi_{i j^{\prime}}=1\right\}$ happens with probability $\left(\frac{J_{s}}{d}\right)\left(1-\frac{J_{s}}{d}\right)$ and event $\left\{\xi_{i j}=0, \xi_{i j^{\prime}}=0\right\}$ happens with probability $\left(1-\frac{J_{s}}{d}\right)^{2}$.

Proposition \ref{THM:Random-k-spatial 1} infers that $\mathbb{E}\|\hat{G}-\bar{G}\|^{2}$ can be directly computed as
\begin{align*}
&\mathbb{E}\|\hat{G}-\bar{G}\|^{2}=\sum_{j, j^{\prime}=1}^{d} \mathbb{E}\left(\hat{G}_{jj^{\prime}}-\bar{G}_{jj^{\prime}}\right)^{2}\notag\\
=&\sum_{j, j^{\prime}=1}^{d} \mathbb{E}\left(\frac{1}{n} \frac{\bar{\beta}^{2}}{T\left(M_{j}\right)T\left(M_{j^{\prime}}\right)}\sum_{i=1}^{n}\left(h_{ij}-\bar{h}_{j}\right)\left(h_{ij^{\prime}}-\bar{h}_{j^{\prime}}\right)-\frac{1}{n}\sum_{i=1}^{n}\left(x_{ij}-\bar{m}_{j}\right)\left(x_{ij^{\prime}}-\bar{m}_{j^{\prime}}\right)\right)^{2}\\
=&\frac{1}{n^{2}}\sum_{j, j^{\prime}=1}^{d} \mathbb{E}\left(\frac{\bar{\beta}^{2}}{T\left(M_{j}\right)T\left(M_{j^{\prime}}\right)}\sum_{i=1}^{n}\left(h_{ij}-\bar{h}_{j}\right)\left(h_{ij^{\prime}}-\bar{h}_{j^{\prime}}\right)\right)^{2}-\frac{1}{n^{2}}\left(\sum_{i=1}^{n}\left(x_{ij}-\bar{m}_{j}\right)\left(x_{ij^{\prime}}-\bar{m}_{j^{\prime}}\right)\right)^{2}
\end{align*}

The following theorem to measure the quality of Random-knots-Spatial estimator,
\begin{theorem}
\label{THM:Random-k-spatial 2}
(Random-knots-Spatial Estimation Error). Under Assumptions 1--5, MSE of estimate $\hat{G}$ produced by the Random-knots-Spatial family is given by
\begin{align}
\label{diff}
\mathbb{E}\|\hat{G}-\bar{G}\|^{2}=\frac{1}{n^{2}}\left(\left(\frac{d}{J_{s}}+c_{1}\right)^{2}-1\right) R_{1}+\frac{1}{n^{2}}\left(\left(1-c_{2}\right)^{2}-1\right) R_{2}
\end{align}
where $R_{1}=\sum_{i=1}^{n}\left\|x_{i}-\bar{m}\right\|^{4}$ and $R_{2}=2 \sum_{i=1}^{n} \sum_{k=i+1}^{n}\left\langle\left(x_{i}-\bar{m}\right)^{2}, \left(x_{k}-\bar{m}\right)^{2}\right\rangle$. The parameters $c_{1}$, $c_{2}$ depend on the choice of $T\left(\cdot\right)$ as
\begin{align*}
&c_{1}=\bar{\beta}^{2} \sum_{r=1}^{n} \frac{J_{s}}{d T(r)^{2}}\binom{n-1}{r-1}\left(\frac{J_{s}}{d}\right)^{r-1}\left(1-\frac{J_{s}}{d}\right)^{n-r}-\frac{d}{J_{s}}\\
&c_{2}=1-\bar{\beta}^{2} \sum_{r=2}^{n} \frac{J_{s}^{2}}{d^{2} T(r)^{2}}\binom{n-2}{r-2}\left(\frac{J_{s}}{d}\right)^{r-2}\left(1-\frac{J_{s}}{d}\right)^{n-r} 
\end{align*}
where $\bar{\beta}$ is defined in (\ref{def:betabar}).

Assumption 6 further guarantees that
$
\left\|\hat{G}-\bar{G}\right\|_{2}=\mathcal{O}_{p}\left(n^{-1/2}\right).
$
\end{theorem}

The result in (\ref{diff}) can be further simplify as
\begin{align*}
\mathbb{E}\|\hat{G}-\bar{G}\|^{2}=\frac{1}{n^{2}}\left(\left(\frac{d}{J_{s}}\right)^{2}+c_{1}^{2}+2c_{1}\frac{d}{J_{s}}-1\right) R_{1}+\frac{1}{n^{2}}\left(c_{2}^{2}-2c_{2}\right) R_{2}
\end{align*}
Note that $R_{1}+R_{2}=\left\|\sum_{i=1}^{n} \left(x_{i}-\bar{m}\right)^{2}\right\|^{2}\geq 0$ and $\left\|\sum_{i=1}^{n} \left(x_{i}-\bar{m}\right)^{2}\right\|^{2} \leq n R_{1}$, $R_{1}\geq 0$, $R_{2}\geq 0$, it follows that $\frac{R_{2}}{R_{1}} \in[0, n-1]$.
MSE of Random-knots-Spatial covariance estimator is guaranteed to be lower than the Random-knots covariance estimator whenever 
$$\left(c_{1}^{2}+2c_{1}\frac{d}{J_{s}}\right)R_{1}<\left(2c_{2}-c_{2}^{2}\right)R_{2}\quad i.e. \quad\frac{R_{2}}{R_{1}}>\frac{c_{1}^{2}+2c_{1}\frac{d}{J_{s}}}{2c_{2}-c_{2}^{2}}.$$ In general, since the MSE
depends on the function $T(\cdot)$ through $c_{1}$ and $c_{2}$, we can find the $T(\cdot)$ that ensures that Random-knots-Spatial estimate is more accurate than Random-knots estimate. And finding optimal $T(\cdot)$ that minimizes MSE as
shown below.

\begin{theorem}
\label{THM:Random-k-spatial 3}
(Random-knots-Spatial minimum MSE) The estimator within the Random-knots-Spatial estimators that minimizes the MSE in (\ref{diff}), can be obtained by setting
\begin{align}
\label{opt}
T^{*}(r)=\left(1+\frac{R_{2}}{R_{1}} \left(\frac{r-1}{n-1}\right)^{2}\right)^{1/2}, \quad  r=1, \ldots, n .
\end{align}
\end{theorem}
Two settings are considered below.

Random-knots-Spatial (Min) is exactly Random-knots estimator. Random-knots estimator reaches the minimum MSE when $R_{2} / R_{1}=0$, which means that $R_{2}=2 \sum_{i=1}^{n} \sum_{k=i+1}^{n}\left\langle\left(x_{i}-\bar{m}\right)^{2}, \left(x_{k}-\bar{m}\right)^{2}\right\rangle=0$ while $R_{1}=\sum_{i=1}^{n}\left\|x_{i}-\bar{m}\right\|^{4}>0$. The optimal MSE occurs when the de-meaned node vectors $\left\{x_{i}-\bar{m}\right\}_{i=1}^{n}$ are orthogonal to de-meaned vector $\left\{x_{k}-\bar{m}\right\}_{k=i+1}^{n}$ and vectors of $n$ nodes are not constant. In this case, $T^{*}(r)=1, \forall r$ in (\ref{opt}) reaching the minimum value. Hence, when all de-meaned vectors are orthogonal, i.e. no correlation between the nodes, the Random-knots covariance estimator is the best estimator. However, when nodes are correlated ($R_{2} / R_{1} \neq 0$), Random-knots estimator is not the best. It is necessary to compute $R_{2} / R_{1}$ based on different node vectors $\left\{x_{i}\right\}_{i=1}^{n}$, and further obtain estimator corresponding to value of $T^{*}(\cdot)$.

Random-knots-Spatial (Max) A special case is that $x_{i}=x_{j}$, $1\leq i, j \leq n$. In this case, $R_{2} / R_{1}=n-1$, and $T^{*}(r)=\left(1+\frac{\left(r-1\right)^{2}}{n-1}\right)^{1/2}$ gives the optimal estimator. Random-knots-Spatial (Max) estimator corresponds to the maximum value of $R_{2} / R_{1}$ and the highest correlation between nodes. Thus, the optimal estimators are very different when the vectors are uncorrelated as opposed to when the vectors are highly correlated.

\cite{Jhun21} has claimed that optimal Random-knots-Spatial mean estimator is obtained when $T^{*}(r)=1+\frac{R_{2}}{R_{1}}\frac{r-1}{n-1}$. The optimal setting of mean estimation cannot guarantee the optimal estimation of covariance function. Therefore, we need to figure out how to set spatial function $T^{*}\left(\cdot\right)$ depending on the specific task. For tasks involving only mean function estimation, such as K-means, the setting proposed by \cite{Jhun21} should be adopted. However, for tasks involving only covariance function estimation, such as PCA, the MSE of covariance statistics set as (\ref{opt}) is smaller. The mean and covariance functions can reflect the distribution characteristics of data. It is significant to estimate the covariance function accurately so as to improve the efficiency of the task and accurate eigenvalues and eigenfunctions can be further obtained. In this way, the extracted features can improve the accuracy of downstream tasks of classification and regression.

In fact, the large number of nodes $n$ and dimension of vector $d$ yield the amount of computation for $R_{1}$ and $R_{2}$. We propose the following Random-knots-Spatial (Avg) as a default setting, and effectiveness is shown in simulation studies.
$$
\tilde{T}(r)=\left(1+\frac{n}{2} \left(\frac{r-1}{n-1}\right)^{2}\right)^{1/2}, \quad  r=1, \ldots, n .
$$

\subsection{Fix-Sparsification}
We retain elements at $J_{s}$ fixed position and set the rest to zero. This method of dimensionality reduction only utilizes values at a fixed number of positions in the vector in each iteration. Although the unbiased estimate of each node vector is obtained after scaled by a factor $\frac{J_{s}}{d}$, it enjoys the following disadvantages: (i) each step can only leverages the subset of size $n\times J_{s}$ of the origin data set of size $n\times d$ where $J_{s}\ll d $, and the subset selected is fixed and will not change along with training steps. This results in a serious loss of information, and the accuracy of the estimates does not increase with the number of training sessions. (ii) The approximate quality of the estimator depends on the selected points, especially if the selected points deviate from the overall distribution, then the MSE of the estimator is difficult to control. (iii) Even if the suitable fixed points for the vector of node $1$ are determined by adding penalty terms or artificially selected methods, there is no guarantee that these fixed points are correct for a different node $2$. 
\begin{figure}[!h]
    \centering
    \includegraphics[width=\linewidth]{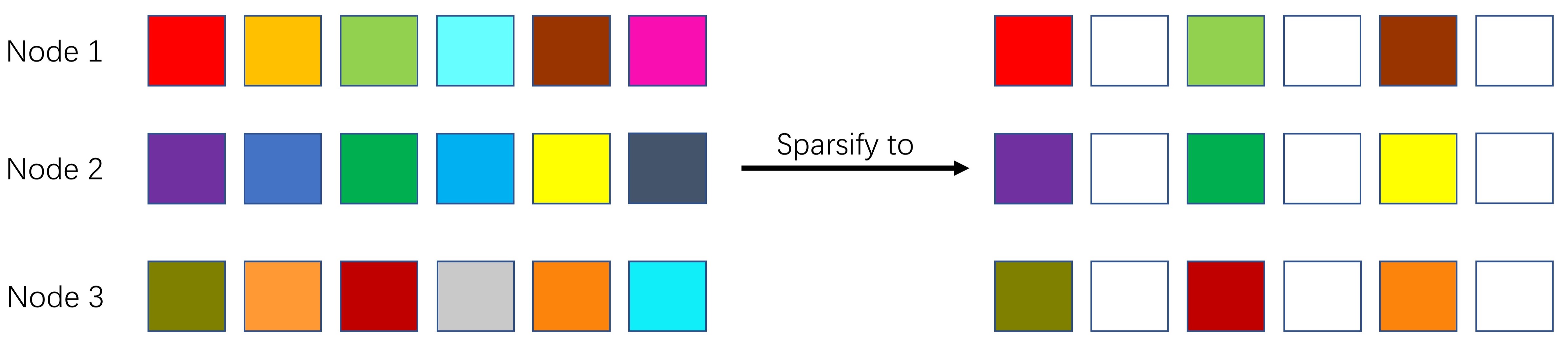}
    \caption{The fix-knots method for $n=3$ nodes, left: origin vectors $\left\{x_{i}\right\}_{i=1}^{3}$ with dimension $d=6$, right: sparsified vectors $\left\{h_{i}\right\}_{i=1}^{3}$ and determine $J_{s}=3$ non-zero values at three fixed positions for each vector.}
\end{figure}

\noindent\textbf{B-spline estimator} B-spline interpolation reduces the loss of information by fitting the values between fixed points. A reasonable alternative to interpolation is to consider the correlation between the elements in the vector, and the closer the points are to each other, the greater the correlation. It is worth mentioning that the choice of basis functions and other fitting methods (polynomial, kernel and wavelet smoothing) do not affect the large-sample theories, according to our proofs. We choose B-spline basis functions because they are more computationally efficient and numerically stable in finite samples compared with other basis functions such as the truncated power series and trigonometric series, which is very
suitable for analyzing large data sets without uniform distribution. (see \cite{Sch07}). 

Denote by $\left\{t_{\ell}\right\}_{\ell=1}^{J_{s}}$ a sequence of equally-spaced points, $t_{\ell}=\ell /\left(J_{s}+1\right), \ell \in\left\{1, \ldots, J_{s}\right\}, 0<t_{1}<\cdots<t_{J_{s}}<1$, called interior knots, which divide the interval $[0,1]$ into $\left(J_{s}+1\right)$ equal subintervals $I_{0}=\left[0, t_{1}\right), I_{\ell}=\left[t_{\ell}, t_{\ell+1}\right)$, $\ell \in\left\{1, \ldots, J_{s}-1\right\}$, $I_{J_{s}}=\left[t_{J_{s}}, 1\right]$. For any positive integer $p$, let $t_{1-p}=\cdots=t_{0}=0$ and $1=t_{J_{s}+1}=\cdots=t_{J_{s}+p}$ be auxiliary knots. Let $\mathcal{S}^{(p-2)}=\mathcal{S}^{(p-2)}[0,1]$ be the polynomial spline space of order $p$ on $I_{\ell}$, $\ell \in\left\{0, \ldots, J_{s}\right\}$, which consists of all $(p-2)$ times continuously differentiable functions on $[0,1]$ that are polynomials of degree $(p-1)$ on subintervals $I_{\ell}$, $\ell \in\left\{0, \ldots, J_{s}\right\}$. We denote by $\left\{B_{\ell, p}(x), 1 \leq \ell \leq J_{s}+p\right\}$ the pth order B-spline basis functions of $\mathcal{S}^{(p-2)}$, hence $\mathcal{S}^{(p-2)}=\left\{\sum_{\ell=1}^{J_{s}+p} \lambda_{\ell, p} B_{\ell, p}(x) \mid \lambda_{\ell, p} \in \mathbb{R}, x \in[0,1]\right\}$.

The $i$ th unknown trajectory $\eta_{i}(x)$ is estimated by using the following formula
\begin{align}
\label{spline tra}
h_{i}(\cdot)=\underset{g(\cdot) \in \mathcal{S}^{\left(p-2\right)}}{\operatorname{argmin}} \sum_{j=1}^{d}\left\{x_{i j}-g\left(x_{j}\right)\right\}^{2}=\sum_{\ell=1}^{J_{s}+p} \hat{\lambda}_{\ell, p, i} B_{\ell, p}(\cdot),\quad 1 \leq i \leq n,
\end{align}
that is the coefficients satisfy
$$
\left(\hat{\lambda}_{1, p, i}, \ldots, \hat{\lambda}_{J_{s}+p, p, i}\right)^{\top}=\underset{\left(\lambda_{1, p}, \ldots, \lambda_{J_{s}+p, p}\right) \in \mathbb{R}^{J_{s}+p}}{\operatorname{argmin}} \sum_{j=1}^{d}\left\{x_{i j}-\sum_{\ell=1}^{J_{s}+p} \lambda_{\ell, p} B_{\ell, p}\left(j / d\right)\right\}^{2} .
$$
\textbf{Decomposition} The design matrix for B-spline regression is
$$
\mathbf{B}=\{\mathbf{B}(1 / d), \ldots, \mathbf{B}(d / d)\}^{\top}=\left(\begin{array}{ccc}
B_{1, p}(1 / d) & \cdots & B_{J_{s}+p, p}(1 / d) \\
\vdots & \cdots & \vdots \\
B_{1, p}(d / d) & \cdots & B_{J_{s}+p, p}(d / d)
\end{array}\right).
$$
Denote by $\mathbf{V}_{n, p}$ the empirical inner product matrix of B-spline basis $\left\{B_{\ell, p}(t)\right\}_{\ell=1}^{J_{s}+p}$, i.e.
$$
\mathbf{V}_{n, p}=\left\{\left\langle B_{\ell, p}, B_{\ell^{\prime}, p}\right\rangle_{d}\right\}_{\ell, \ell^{\prime}=1}^{J_{s}+p}=d^{-1} \mathbf{B}^{\top} \mathbf{B} .
$$
The spline estimator $h_{i}(\cdot)$ allows representation $h_{i}(\cdot)=d^{-1} \mathbf{B}(\cdot)^{\top} \mathbf{V}_{n, p}^{-1} \mathbf{B}^{\top}x_{i}=\hat{m}(\cdot)+\hat{Z}_{i}(\cdot),
$
where
\begin{align*}
&\hat{m}(\cdot)=d^{-1} \mathbf{B}(\cdot)^{\top} \mathbf{V}_{n, p}^{-1} \mathbf{B}^{\top}m,\\
&\hat{Z}_{i}(\cdot)=d^{-1} \mathbf{B}(\cdot)^{\top} \mathbf{V}_{n, p}^{-1} \mathbf{B}^{\top} Z_{i}.
\end{align*}
The mean and covariance is obtained by plugging in the newly defined $\left\{h_{i}\right\}_{i=1}^{n}$ in (\ref{Random-k mean}) and (\ref{Random-k cov}).
\begin{figure}[!h]
    \centering
    \includegraphics[width=\linewidth]{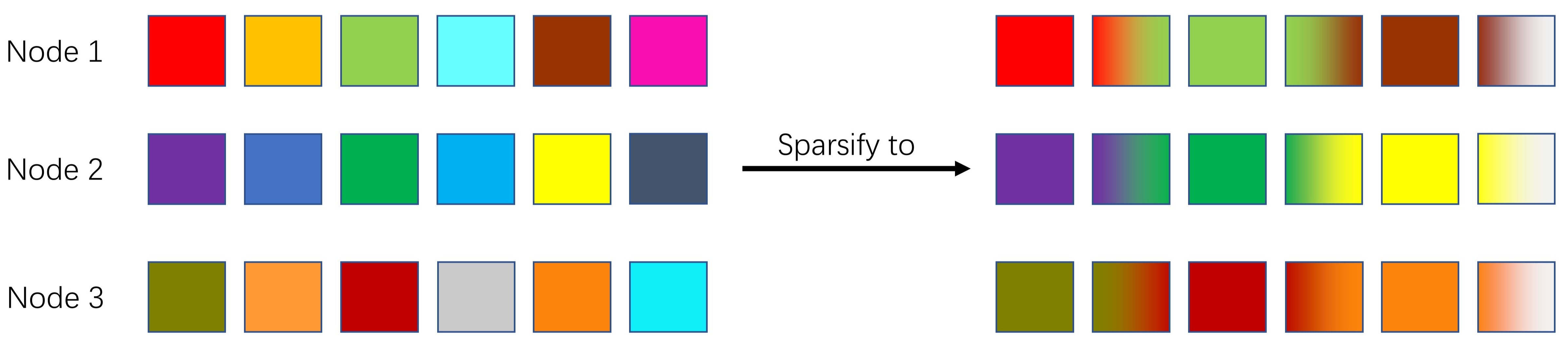}
    \caption{The B-spline interpolation method for $n=3$ nodes, left: origin vectors $\left\{x_{i}\right\}_{i=1}^{3}$ with dimension $d=6$, right: sparsified vectors $\left\{h_{i}\right\}_{i=1}^{3}$ and determine $J_{s}=3$ non-zero values at three fixed positions for each vector, the remaining positions are replaced by B-spline interpolation.}
\end{figure}

The next theorem states the convergence rate of the B-spline estimators. 
\begin{theorem}
\label{THM:spline}
(B-spline mean/covariance Estimation Error). Under Assumptions 1--6, the B-spline estimator $\hat{m}(\cdot)$ is asymptotically equivalent to $\bar{m}(\cdot)$ up to order $n^{1 / 2}$ and the same thing is true for covariance, i.e., as $n \rightarrow \infty$,
\begin{align}
&\|\hat{m}-\bar{m}\|_{\infty}=\sup _{t \in[0,1]}|\hat{m}(t)-\bar{m}(t)|={\scriptstyle{\mathcal{O}}}_{a.s.}\left(n^{-1 / 2}\right)\label{THM:splinemean}\\
&\|\hat{G}-\bar{G}\|_{\infty}=\sup _{t, t^{\prime} \in[0,1]}\left|\hat{G}\left(t, t^{\prime}\right)-\bar{G}\left(t, t^{\prime}\right)\right|=\mathcal{O}_{p}\left(n^{-1 / 2}\right).\label{THM:splinecov}
\end{align}
\end{theorem}

\textbf{Bspline-Spatial estimators} We replace $\left\{h_{ij}\right\}_{i,j=1}^{n,d}$ in (\ref{Random-k spatial mean}), (\ref{Random-k spatial cov}) by their corresponding B-spline estimators in (\ref{spline tra}). The estimators not only takes into account the correlation between nodes, but also the correlation of vectors within a single node, which can minimize the information loss of original data.

\begin{theorem}
\label{THM:spline-spatial}
(Bspline-Spatial mean/covariance Estimation Error).
Under Assumptions 1--6, the Bspline-Spatial estimator $\hat{m}(\cdot)$ is asymptotically equivalent to $\bar{m}(\cdot)$ up to order $n^{1 / 2}$ and the same thing is true for covariance, i.e., as $n \rightarrow \infty$
\begin{align*}
&\|\hat{m}-\bar{m}\|_{\infty}=\sup _{t \in[0,1]}|\hat{m}(t)-\bar{m}(t)|={\scriptstyle{\mathcal{O}}}_{p}\left(n^{-1 / 2}\right)\\
&\|\hat{G}-\bar{G}\|_{\infty}=\sup _{t, t^{\prime} \in[0,1]}\left|\hat{G}\left(t, t^{\prime}\right)-\bar{G}\left(t, t^{\prime}\right)\right|=\mathcal{O}_{p}\left(n^{-1 / 2}\right).
\end{align*}
\end{theorem}

\subsection{Convergence of Principal Component}
\label{SEC:fpc}
The estimates of eigenfunctions and eigenvalues is obtained by solving the following eigenequations,
\begin{align}
\label{int}
\int_{0}^{1}\hat{G}\left(x, x^{\prime}\right) \hat{\psi}_{k}\left(x^{\prime}\right) d x^{\prime}=\hat{\lambda}_{k} \hat{\psi}_{k}\left(x\right).
\end{align}
The following theorem obtain the consistency of $\hat{\lambda}_{k}$ in $\left(\ref{int}\right)$ for $\lambda_{k}$. By choosing $\hat{\phi}_{k}$ appropriately when $\lambda_{k}$ is of multiplicity $1$, $\hat{\psi}_{k}$ converges to $\psi_{k}$ uniformly on the bounded interval $\left[0,1\right]$.
\begin{theorem}
\label{THM:FPC} 
As $n\rightarrow\infty$, for $k\in\mathbb{N}$,
\begin{align}
&\left|\hat{\lambda}_{k}-\lambda_{k}\right| =\mathcal{O}_{p}\left(n^{-1/2}\right)\label{addone}\\
&\left\|\hat{\psi}_{k}-\psi_{k}\right\|_{2} =\mathcal{O}_{p}\left(n^{-1/2}\right)\label{addtwo}\\
&\left\|\hat{\psi}_{k}-\psi_{k}\right\|_{\infty}=\mathcal{O}_{p}\left(n^{-1/2}\right).\label{addthree}
\end{align}
\end{theorem}

Although, in theory, the Karhunen-Loéve representation
of the covariance function consists of an infinite number of
terms. In applications, it is typical to truncate the spectral
decomposition to an integer chosen so as to account for some
predetermined proportion of the variance. One can select the
number of principal components using the Akaike information criterion (AIC; \cite{Yao05}) or Bayesian information
criterion (BIC; \cite{Li13}).

According to $\int \left\{x_{i}
\left(t\right)-m\left(t\right) \right\} \phi_{k}\left(t\right) dx=\lambda _{k}\xi _{ik}$, one obtains
\begin{align*}
\xi_{ik}=\lambda_{k}^{-1/2}\int \left\{x_{i}
\left(t\right)-m\left(t\right) \right\} \psi_{k}\left(t\right) dx.
\end{align*}
Similarly, $\hat{\xi}_{ik}=\hat{\lambda}_{k}^{-1/2}\int \left\{h_{i}
\left(t\right)-\hat{m}\left(t\right) \right\} \hat{\psi}_{k}\left(t\right) dt$. 
The following theroem provides the convergence rate of FPC scores:
\begin{corollary}
\label{THM:FPCscore} 
As $n\rightarrow\infty$
\begin{align*}
\max_{1\leq i\leq n}\left\|\hat{\xi}_{ik}-\xi_{ik}\right\|=\mathcal{O}_{p}\left(n^{-1/2}\right).
\end{align*}
\end{corollary}

As discussed in \cite{Dau82}, for the eigenvalue $\left\{\lambda_{k}\right\}_{k=1}^{\kappa}$ of covariance $G\left(\cdot,\cdot\right)$ with multiplicity greater than $1$, the orthonormal basis of the eigenmanifold corresponding to $\left\{\lambda_{k}\right\}_{k=1}^{\kappa}$ may be 
obtained by rotation, that is the specified eigenvector minus $-1$. Therefore, after calculating the eigendecomposition, the unique form of the eigenfunction should be determined by minimizing the estimation error. The proposed convergence property are for 
each corresponding eigenvectors that have been rotated.

We measure estimation error through the following loss, 
\begin{align*}
L(\hat{\phi}_{k}, \phi_{k}) & \equiv \frac{1}{2} \min _{s \in\{+1,-1\}}\|\hat{\phi}_{k}-s \phi_{k}\|^{2} \\
&=1-|\langle\hat{\phi}_{k}, \phi_{k}\rangle|
\end{align*}
for $\hat{\phi}_{k}, \phi_{k} \in S^{\kappa-1}=\left\{\mathbf{v} \in \mathbb{R}^{\kappa}: \left\|\mathbf{v}\right\|=1\right\}$.
Notice the minimization over the sign $s \in\{+1,-1\}$. This is required because the estimated principal components $\left\{\hat{\phi}_{k}\right\}_{k=1}^{\kappa}$ are only identifiable up to a sign. Analogous results can obtained for alternate loss functions such as the projection distance:
$$
L_{p}(\hat{\phi}_{k}, \phi_{k}) = \frac{1}{\sqrt{2}}\left\|\hat{\phi}_{k}\hat{\phi}_{k}^{\top}-\phi_{k}\phi_{k}^{\top}\right\|_{2}=\sqrt{1-\langle\hat{\phi}_{k}, \phi_{k}\rangle^{2}} .
$$

\section{Implementation}
\label{SEC:imp}
In this section, we state some issues that need to be addressed when constructing
estimation for covariance function and related eigenvalues and eigenvectors.

\subsection{Knots selection}
\label{AIC}
The number of knots is often treated as an unknown tuning parameters, and the
fitting results can be sensitive to it. Since the in-sample fitting errors cannot gauge the
prediction accuracy of the fitted function, we select a criterion function that attempts to
measure the out-of-sample performance of the fitted model. Minimizing the Akaike information criterion (AIC) is one computationally efficient approach to selecting
smoothing parameters that also has good theoretical properties.

As the vector dimension of each node increases, the number of knots needs to be selected will increase accordingly, so as to comprehensively describe the change trend of the vector. If $d=1$, it's enough to select $J_{s}=1$, while not enough if $d=100$. Hence we set the candidate pool for $J_{s}$ is all the integers between $1$ and $J_{s^{*}}$, where $J_{s}^{*}=\min \{10,\lfloor d / 2\rfloor\}$. Specifically, given any data set $\left(x_{i j}, j / d\right)_{i=1, j=1}^{n, d}$ from model (\ref{model}), denote the estimator for the $j$-th response $x_{i j}$ by $h_{i j}\left(J\right)$, for $j=1, \ldots, d$. The trajectory estimate $h_{i}$ depends on the knot selection sequence, as the sparsified vector with a lot of zero components for Random-knots (Random-knots-Spatial) estimators and equispaced spline interpolation vector for B-spline (Bspline-Spatial) estimators. Then, $\hat{J_{s, i}}$ for the $i$-th curve is the one minimizing the AIC value
$$
\hat{J_{s, i}}=\underset{J \in\left[1, J_{s^{*}}\right]}{\arg \min } \operatorname{AIC}\left(J\right), \quad i=1, \ldots, n
$$
where $\operatorname{AIC}\left(J\right)=\log (\mathrm{RSS} / d)+2\left(J+p\right) / d$, with the residual sum of squares $\mathrm{RSS}=\sum_{j=1}^{d}\left\{x_{i j}-h_{i j}\left(J\right)\right\}^{2}$. Then, $\hat{J}_{s}$ is set as the median of $\left\{\hat{J_{s, i}}\right\}_{i=1}^{n}$. The value of $p$ depends on what method is used to get the sparsified trajectory estimate $h_{i}$ where we set $p = 4$ for cubic spline, $p=2$ for linear spline and $p=0$ for estimates that do not involve interpolation.

The trajectory estimator $h_{i}(j/d)$ is obtained by using the selected number of knots $\hat{J}_{s}$, and the mean estimator $\hat{m}(\cdot)$ is computed from (\ref{Random-k mean}), the covariance estimator $\hat{G}(\cdot, \cdot)$ is computed from (\ref{Random-k cov}).

\subsection{Truncation}
One truncates the spectral decomposition at an integer $\kappa$ which is selected according to the standard criteria called "pseudo-AIC" in \cite{Mu08} that the number of eigenvalues can explain $95 \%$ of the variation in the data. That is to say,
$$
\kappa=\operatorname{argmin}_{1-p \leq \ell \leq J_{s}}\left\{\sum_{k=1-p}^{\ell} \hat{\lambda}_{k} / \sum_{k=1-p}^{J_{s}} \hat{\lambda}_{k}>0.95\right\}
$$
where $\left\{\hat{\lambda}_{k}\right\}_{k=1-p}^{J_{s}}$ are all nonnegative eigenvalues estimated in FPC analysis. This simple method of counting the percentage of variation explained can be used to choose the number of principal components.

\section{Simulation}
\label{SEC:sim}
We conduct simulation studies to illustrate the
finite-sample performance of the proposed methods.

\subsection{Accuracy of covariance estimator}

The data are generated from the following model: 
$$x_{i j}=m(j / d)+\sum_{k=1}^{\infty} \xi_{i k} \phi_{k}(j / d), \quad j \in\{1, \ldots, d\},\quad i \in\{1, \ldots, n\}$$
where $m(t)=\sin \{2 \pi(t-1 / 2)\}$, $\phi_{k}(t)=\sqrt{\lambda_{k}} \psi_{k}(t)$ with $\lambda_{k}=(1 / 4)^{[k / 2]}$, $\psi_{2 k-1}(t)=\sqrt{2} \cos (2 k \pi t)$, $\psi_{2 k}(t)=\sqrt{2} \sin (2 k \pi t)$, $ k \geq 1$. $\xi_{ik}$ are i.i.d. standardized normal distribution. The infinite series $G\left(t, t^{\prime}\right)=\sum_{k=1}^{\infty} \phi_{k}(t) \phi_{k}\left(t^{\prime}\right)$ is well approximated by finite sum $G\left(t, t^{\prime}\right)=\sum_{k=1}^{k_{0}} \phi_{k}(t) \phi_{k}\left(t^{\prime}\right)$ where $k_{0}=1000$, as the fraction of variance explained (FVE) criteria, $\mathrm{FVE}=\sum_{k=1}^{1000} \lambda_{k} / \sum_{k=1}^{\infty} \lambda_{k}>1-10^{-10}$, see \cite{Yao05}. We fix the number of nodes $n$ is $200$ and the value range of dimension of vector $d$ is set a point every $25$ points between $50$ and $400$. Similarly, $n$ is set to vary equally between $50$ and $400$ with fixed $d=200$. Each simulation is repeated $1000$ times. Throughout this section, the covariance function is obtained by $p=4$ for B-spline (Bspline-Spatial) estimator and $p=0$ for Random-knots (Random-knots-Spatial) estimator, with the number of knots selected using the AIC given in Section \ref{AIC}.

\begin{figure}[htbp]
\centering

\begin{subfigure}{0.45\textwidth}
  \centering
  \includegraphics[width=\linewidth]{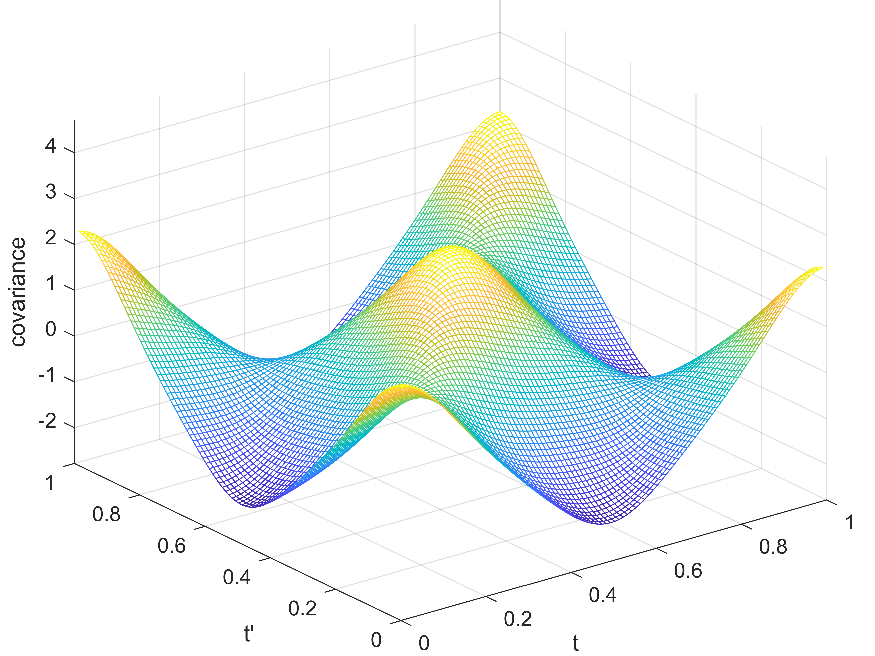}
  \caption{$G\left(t, t^{\prime}\right)$}
\end{subfigure}
\begin{subfigure}{0.45\textwidth}
  \centering
  \includegraphics[width=\linewidth]{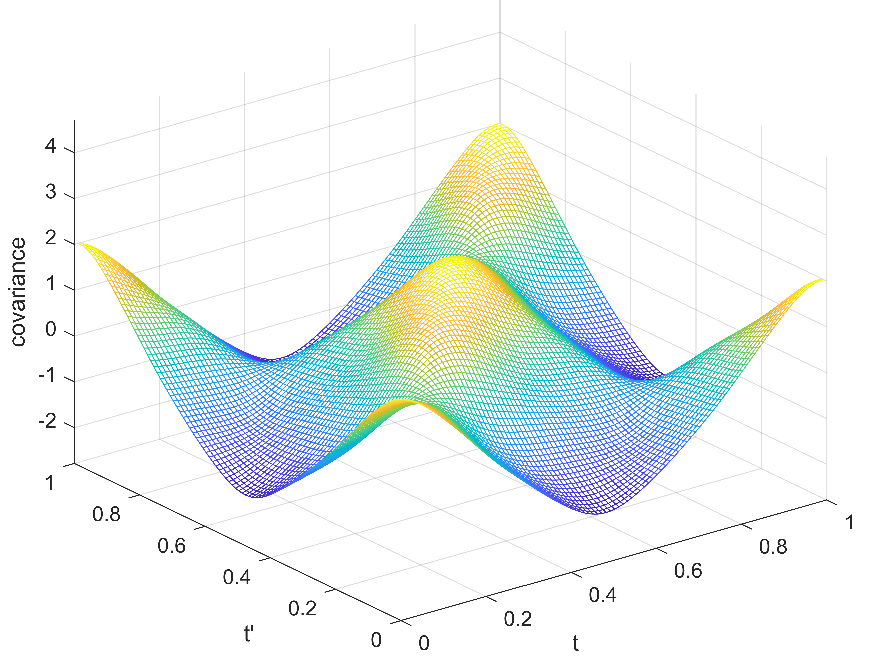}
  \caption{$\bar{G}\left(t, t^{\prime}\right)$}
\end{subfigure}

\begin{subfigure}{0.45\textwidth}
  \centering
  \includegraphics[width=\linewidth]{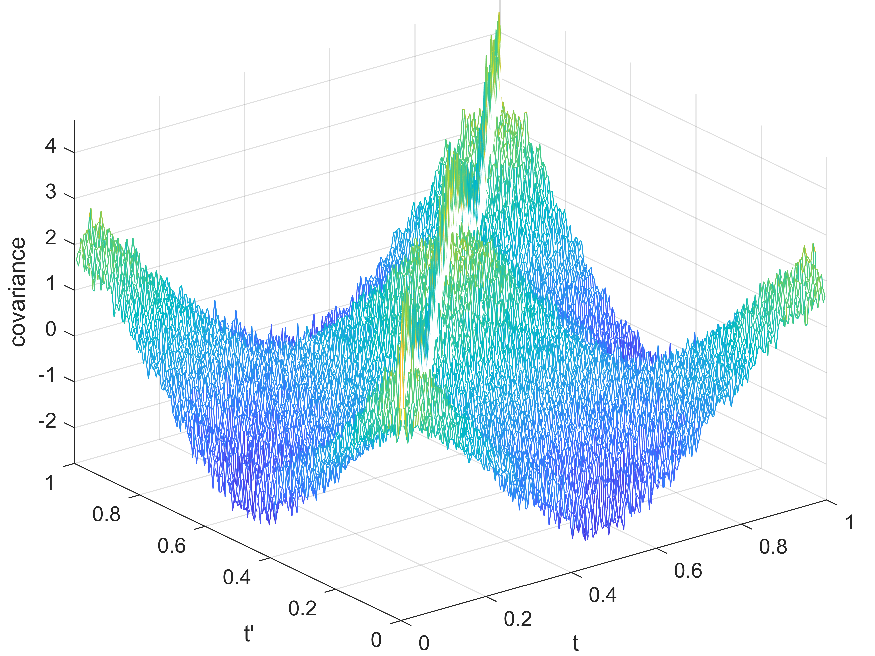}
  \caption{Random-knots $\hat{G}\left(t, t^{\prime}\right)$}
\end{subfigure}
\begin{subfigure}{0.45\textwidth}
  \centering
  \includegraphics[width=\linewidth]{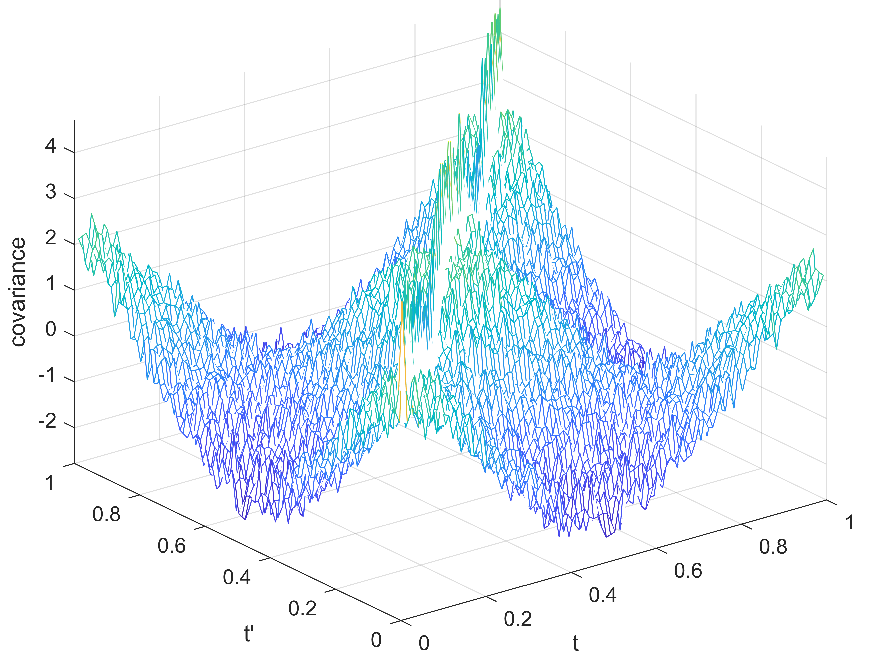}
  \caption{Random-knots-Spatial $\hat{G}\left(t, t^{\prime}\right)$}
\end{subfigure}

\begin{subfigure}{0.45\textwidth}
  \centering
  \includegraphics[width=\linewidth]{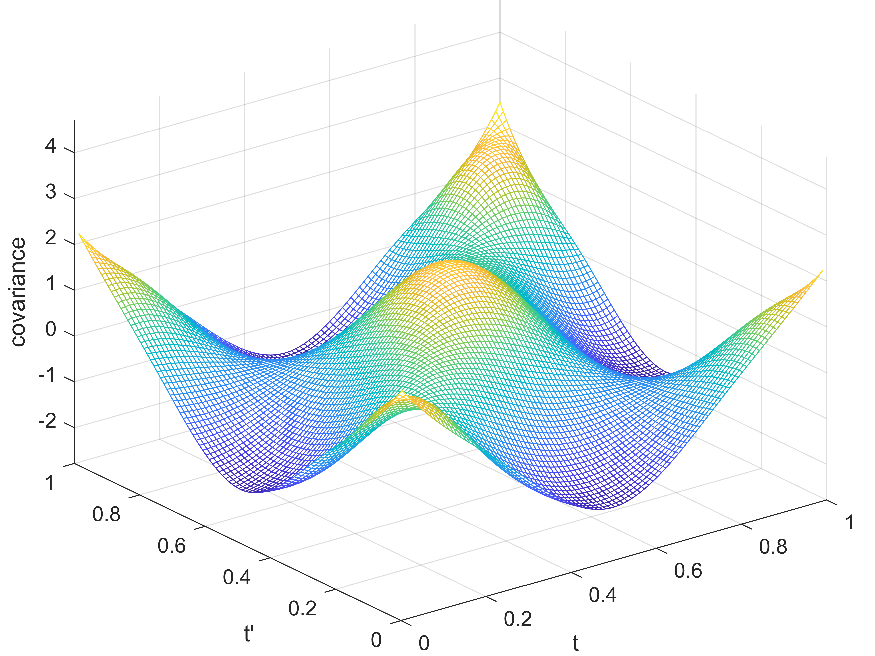}
  \caption{B-spline $\hat{G}\left(t, t^{\prime}\right)$}
\end{subfigure}
\begin{subfigure}{0.45\textwidth}
  \centering
  \includegraphics[width=\linewidth]{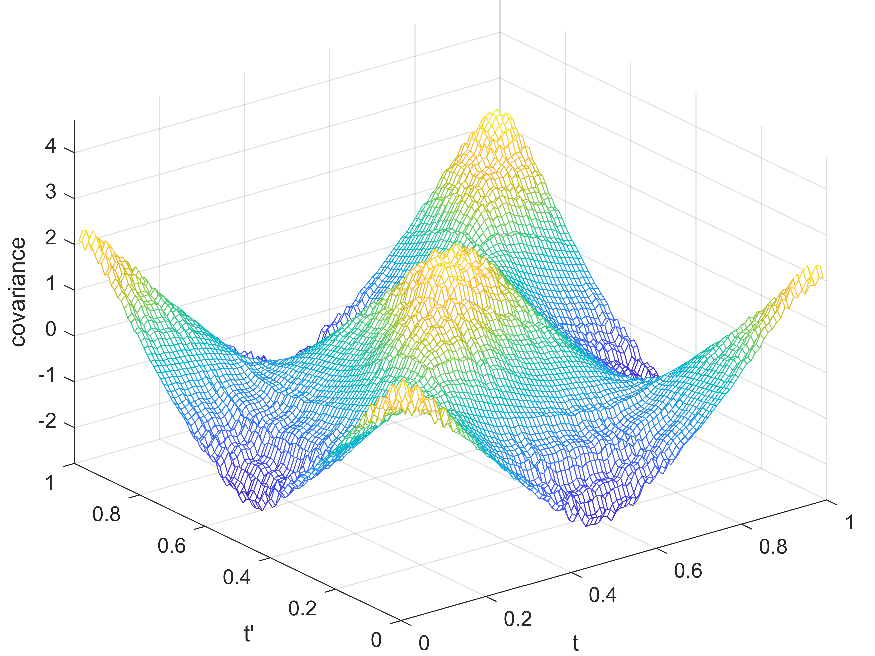}
  \caption{Bspline-Spatial $\hat{G}\left(t, t^{\prime}\right)$}
\end{subfigure}

\caption{Plots of true covariance, averaged covariance and four different covariance estimators.}
\label{plot_G}
\end{figure}

We examine the accuracy of the proposed two-stage estimation procedure. The average mean squared error (AMSE) is computed to assess the performance of the covariance estimators $\hat{G}(\cdot, \cdot)$ and $\bar{G}(\cdot, \cdot)$ defined in (\ref{Random-k cov}) and (\ref{plot_G}), respectively. The AMSE of $\hat{G}(\cdot, \cdot)$ is defined as
\begin{align*}
\operatorname{AMSE}\left(\hat{G}\right)=\frac{1}{1000 d^{2}} \sum_{s=1}^{1000} \sum_{j, j^{\prime}=1}^{d}\left\{\hat{G}_{s}\left(j / d, j^{\prime} / d\right)-\bar{G}\left(j / d, j^{\prime} / d\right)\right\}^{2}, 
\end{align*}
where $\hat{G}_{s}$ represents the values of the $s$-th replication of sample covariance $\hat{G}(\cdot, \cdot)$ in(\ref{Random-k cov}) and $\bar{G}_{s}$ represents the values of the $s$-th replication of $\bar{G}(\cdot, \cdot)$ in(\ref{plot_G}). Furthermore, given the true covariance function $G\left(\cdot, \cdot\right)$, we obtain the MSE between $\bar{G}\left(\cdot, \cdot\right)$ and $G\left(\cdot, \cdot\right)$ as
\begin{align*}
\operatorname{AMSE}\left(\bar{G}\right)=\frac{1}{1000 d^{2}} \sum_{s=1}^{1000} \sum_{j, j^{\prime}=1}^{d}\left\{\bar{G}_{s}\left(j / d, j^{\prime} / d\right)-G\left(j / d, j^{\prime} / d\right)\right\}^{2}.
\end{align*}

\begin{figure}[htbp]
\centering
\includegraphics[width=\linewidth,height=5cm]{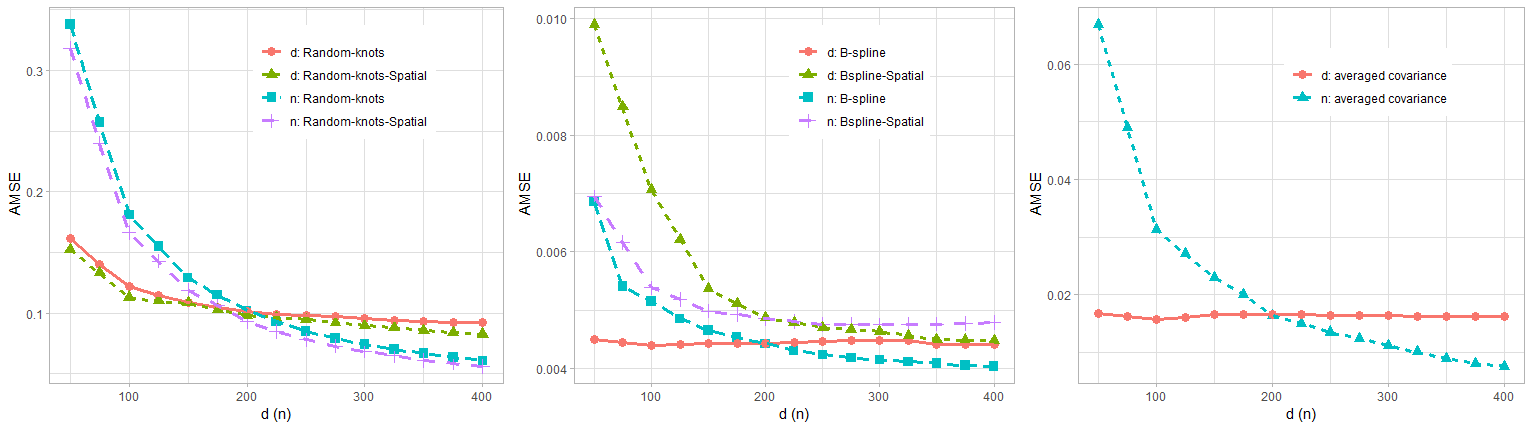}
\caption{Left, middle: $\operatorname{AMSE}\left(\hat{G}\right)$ as a function of $d$ or $n$. Right: $\operatorname{AMSE}\left(\bar{G}\right)$ as a function of $d$ or $n$.}
\label{AMSE_G}
\end{figure}

In the left panel of Figure \ref{AMSE_G}, ``d: Random-knots'' stands for the curve of $\operatorname{AMSE}\left(\hat{G}\right)$ as $d$ changes where $\hat{G}$ is the Random-knots estimator of covariance function.  ``n: Random-knots-Spatial'' stands for the curve of $\operatorname{AMSE}\left(\hat{G}\right)$ as $n$ changes where $\hat{G}$ is the Random-knots-Spatial covariance estimator. The other two curves in left panel are defined similarly. The $\operatorname{AMSE}(\hat{G})$ is getting smaller when $n$ is increasing, consistent with Theorems \ref{Random-k} and \ref{THM:Random-k-spatial 2} that the Random-knots and Random-knots-Spatial estimators $\hat{G}$ converge to $\bar{G}$ at a rate of $\mathcal{O}_{p}\left(n^{-1/2}\right)$. The $\operatorname{AMSE}(\hat{G})$ also shows a slow downward trend with the increase of $d$, mainly because the number of selected nodes $Js$ increases with the increase of $d$, which affects the accuracy of the covariance estimator. By setting $T\left(\cdot\right)$ the Random-knots-Spatial (Avg), the AMSE of estimator that takes spatial factor into account is generally lower than estimator that does not. 

The middle panel shows that the changing pattern of AMSEs of B-spline and Bspline-Spatial covariance estimators are similar as in left panel confirming Theorems \ref{THM:spline} and \ref{THM:spline-spatial}. The accuracy of the estimator obtained by spline interpolation is significantly improved on the whole. The right panel stands for the curve of $\operatorname{AMSE}\left(\bar{G}\right)$ as $d$ and $n$ change. The simulation result confirms the fact that the convergence rate of $\bar{G}$ to $G$ is $\mathcal{O}_{p}\left(n^{-1/2}\right)$.

Figure \ref{plot_G} shows that all four proposed covariance estimator can well describe the shape of $G\left(t, t^{\prime}\right)$ for $t, t^{\prime}\in[0, 1]$. The deviation of Random-knots and Random-knots-Spatial estimators is large on the diagonal $t=t^{\prime}$ and the smoothness of the surfaces are poor. The accuracy of B-spline estimator is significantly improved, with only slight deviation at the boundary points. On the basis of Random-knots estimator, Random-knots-Spatial estimator further considers the spatial factor, so that the estimation accuracy of boundary points is improved but the smoothness of the surface is sacrificed to some extent.

\subsection{Accuarcy of principle components}
\label{SEC:app}
The AMSE of eigenvalue $\hat{\lambda}_{k}$ 's and the eigenfunction $\hat{\phi}_{k}$ 's are defined as
\begin{align*}
&\operatorname{AMSE}(\hat{\lambda})=\frac{1}{1000 \kappa} \sum_{s=1}^{1000} \sum_{k=1}^{\kappa}\left(\hat{\lambda}_{k s}-\lambda_{k}\right)^{2},\\ &\operatorname{AMSE}(\hat{\phi})=\frac{1}{1000 d\kappa} \sum_{s=1}^{1000} \sum_{j=1}^{d} \sum_{k=1}^{\kappa}\left\{\left(\hat{\phi}_{k s}-\phi_{k}\right)(j / d)\right\}^{2},
\end{align*}
where $\hat{\lambda}_{k s}, \hat{\phi}_{k s}$ represent the values of the $s$-th replication of $\hat{\lambda}_{k}$, $\hat{\phi}_{k}$ in (\ref{int}).

\begin{figure}[htbp]
\centering
\includegraphics[width=\linewidth,height=8cm]{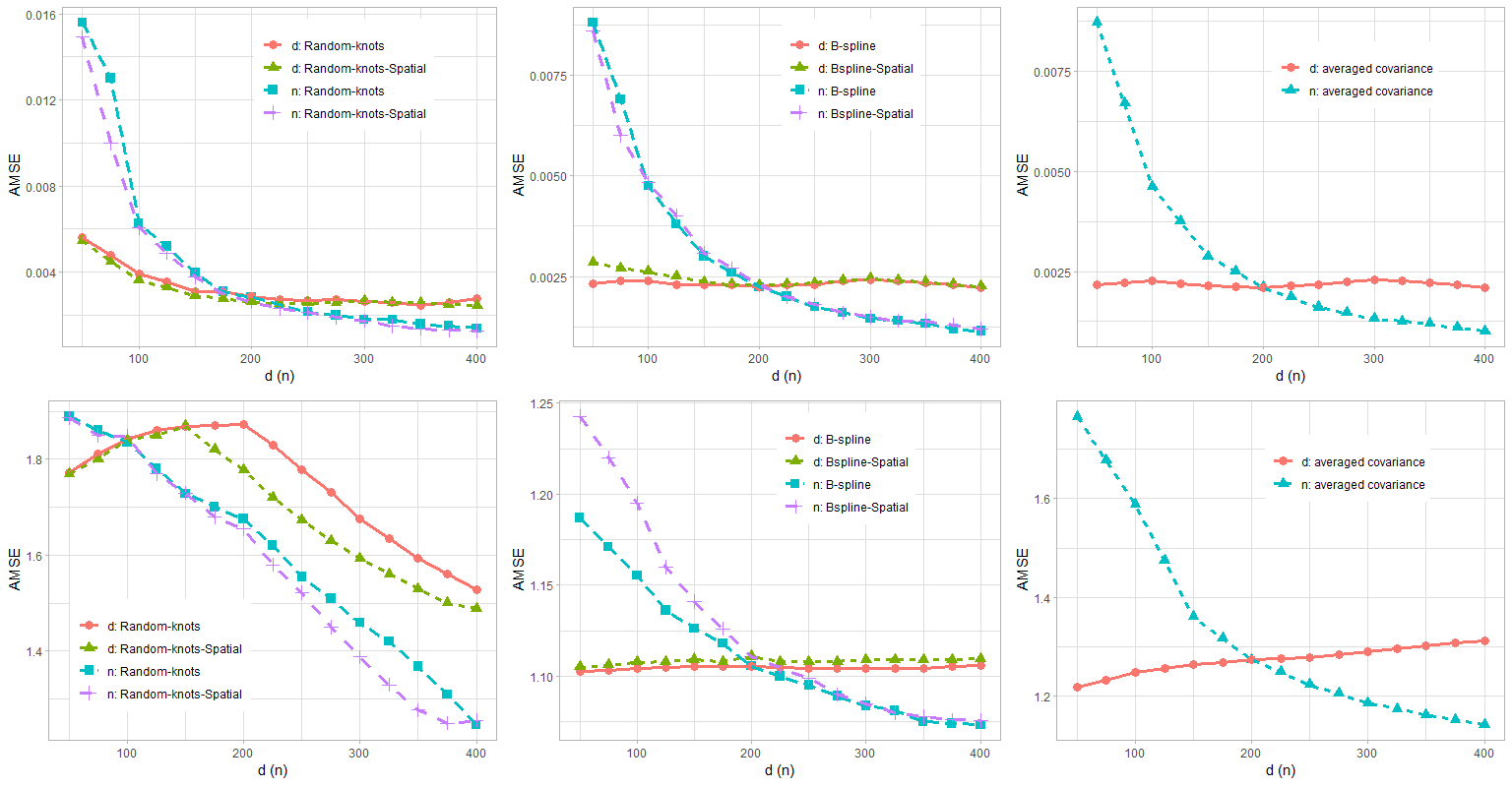}
\caption{Row 1: $\operatorname{AMSE}\left(\hat{\lambda}\right)$ as a function of $d$ or $n$. Row 2: $\operatorname{AMSE}\left(\hat{\phi}\right)$ as a function of $d$ or $n$.}
\label{AMSE_lambda}
\end{figure}

The first row of Figure \ref{AMSE_lambda} reveals that $\operatorname{AMSE}\left(\hat{\lambda}\right)$ decreases with the increase of $n$, but the change with $d$ is small, which is in accordance with Theorem \ref{THM:FPC}. $\hat{\lambda}_{k}$ is the eigenvalue of four different covariance estimators and averaged sample covariance. The second row of Figure \ref{AMSE_lambda} confirms that $\operatorname{AMSE}\left(\hat{\phi}\right)$ exhibits the same regularity as $\operatorname{AMSE}\left(\hat{\lambda}\right)$. Whether or not to consider spatial factors has greater impact on the accuracy of eigenvector than eigenvalue.

\begin{figure}[htbp]
\centering
\includegraphics[width=\linewidth,height=8cm]{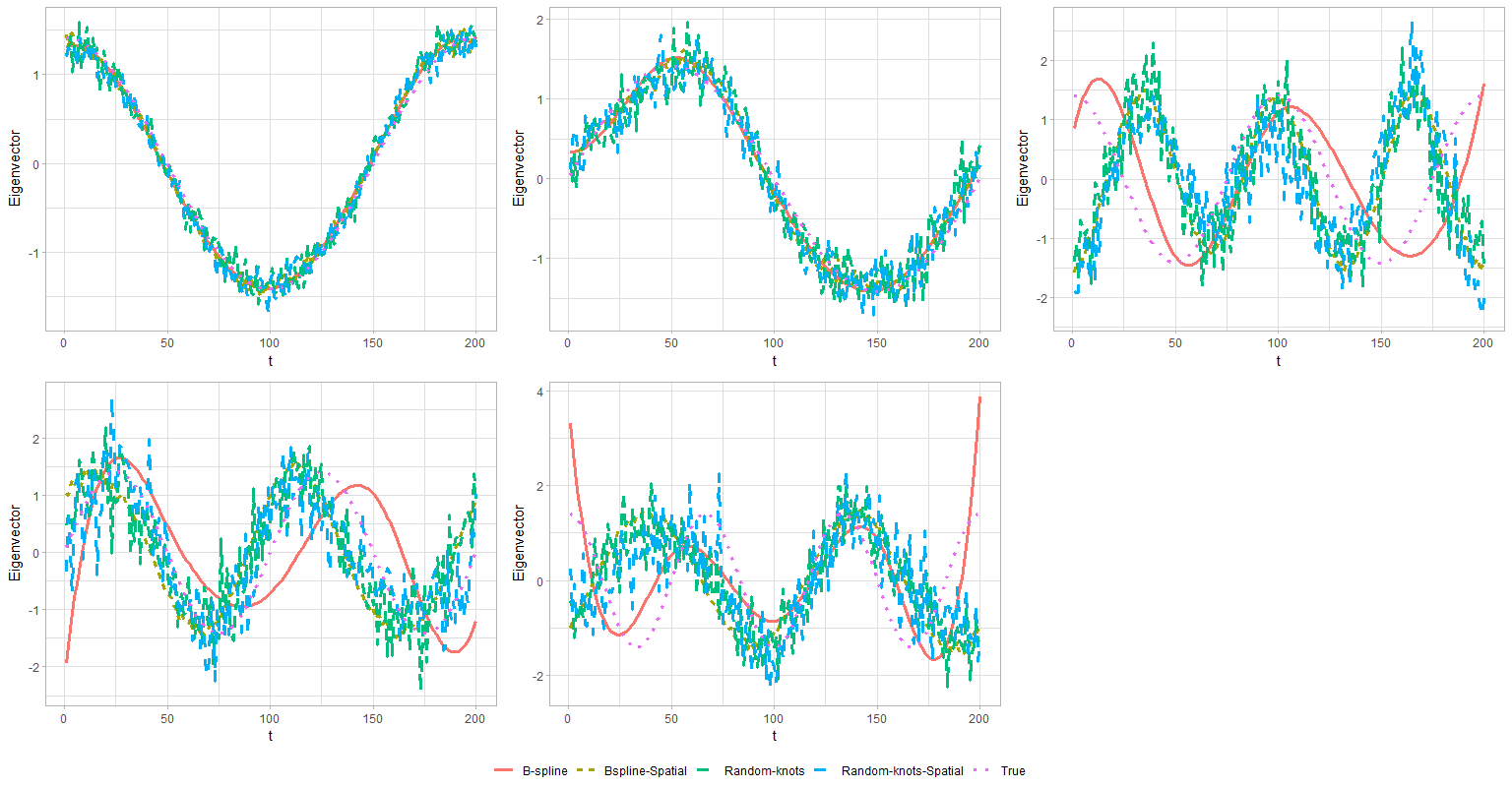}
\caption{Plots of the $k=1,\ldots,5$ eigenfunctions of true covariance and four covariance estimators.}
\label{plotphi}
\end{figure}

In order to visualize the specific form of principle components. Figure \ref{plotphi} illustrates the first five eigenfunctions which
account for $68.2\%$, $17.0\%$, $4.3\%$, $4.3\%$, $4.0\%$ of the total variation. The first figure shows a large difference between the early, middle and late stages of the curve, depicting
the trend term of the covariance function. The other four graphs have great fluctuations
through the whole curves, and the frequency of fluctuations increases with the increase
of $k$. They describe the overall fluctuation characteristics of the covariance function.

\section{Application}
Pre-trained language models
are indeed very useful in clustering sentence representations by domains in an unsupervised manner,
we now seek to harness this property for a downstream task -- domain data selection for machine
translation. 

To evaluate the unsupervised domain clustering we use the multi-domain corpus proposed in \cite{Koe17},
as it was recently adopted for domain adaptation research in NMT (\cite{Hu19}, \cite{Mu19}, \cite{Dou19}, \cite{Dou192}, \cite{Aha20}). The dataset includes parallel text in German and English from five diverse
domains: subtitles
, medical text, legal text, translations of the Koran, and IT-related
text, available via OPUS (\cite{Aul19}). Data split as discussed in \cite{Aha20}.

We encode multi-domain data at the sentence level
into vector representations. For MLM-based models we use BERT \cite{Dev19}, DistilBERT \cite{DBERT19} and
RoBERTa \cite{RO19} (in both the base and
large versions). For autoregressive models we use
GPT-2 \cite{GPT18} and XLNet \cite{XLN19}. In all cases we use the implementations from the HuggingFace Transformers toolkit
\cite{Wolf19}. We also evaluated a simple baseline using representations from word2vec \cite{Mik13},
where we average-pooled the word vectors for the
tokens that were present in the model vocabulary.

We then cluster these
vector representations for each model using a Gaussian Mixture Model (GMM) with $k=5, 10, 15$ pre-defined
clusters. To accelerate the clustering process and
enable visualization we also experiment with performing dimensionality reduction with PCA over
the sentence vectors before clustering them. We experiment with  to test how adding
flexibility would improve the domain clustering
accuracy.

We used $2000$
distinct sentences from each domain. To evaluate
whether the resulting clusters indeed capture the
domains the data was drawn from we measure the
clustering purity, which is a well-known metric
for evaluating clustering (\cite{Sch08}).
To measure the clustering purity, we assign each
unsupervised cluster with the most common true
domain in the sentences assigned to that cluster,
and then compute the accuracy according to this
majority-based cluster-domain assignment. 

Tables \ref{tab1}--\ref{tab3} report unsupervised domain clustering as measured by purity for the different models using original data without sparsification. ``-'' represents result of modeling directly using the original vector and ``PCA'' result of data modeling after dimensionality reduction using PCA. ``RK''(``RK-Spat'') refers to the Random-knots (Random-knots-Spatial) covariance function to calculate eigenvalues and eigenvectors in the process of PCA dimension reduction and ``BS'' (``BS-Spat'') the B-spline (Bspline-Spatial) covariance function. Eight pre-training models are listed in Table \ref{tab:pre}.

\begin{table}[t]
\caption{Serial number and corresponding pre-training model for representations.}
\label{tab:pre}
\begin{center}
\begin{tabular}{clcl}
\toprule
\bf Number & \bf Model        & \bf Number & \bf Model        \\
\midrule
1          & word2vec         & 5          & RoBERTa-base     \\
2          & BERT-base        & 6          & RoBERTa-large    \\
3          & BERT-large       & 7          & GPT-2            \\
4          & DistilBERT       & 8          & XLNet            \\
\bottomrule
\end{tabular}
\end{center}
\end{table}

The
MLM-based models dominated in all cases over
word2vec and the auto-regressive models. This
may be explained by the fact that the MLM-based
models use the entire sentence context when generating the representations for each token, while
the auto-regressive models only use the past context, and word2vec uses a limited window context. 

Using PCA improved performance in most cases and especially for the auto-regressive models, although the results for the MLMs remain high in
both cases, suggesting that these models encode
the information very differently. The four methods we proposed to sparsify the covariance matrix do not sacrifice the purity of the domain-clustering task on the basis of improving the computational speed, and even slightly improve in some cases, reflecting our sparsification methods are efficient and effective.

\begin{table}[t]
\centering
\caption{Purity for the different models using original data without sparsification.}
\label{tab1}
\begin{tabular}{llcccccccc}
\hline\hline
      &        & 1     & 2     & 3     & 4     & 5     & 6     & 7     & 8     \\ \hline
      & -      & 45.93 & 85.81 & 72.25 & 69.99 & 64.91 & 69.84 & 37.82 & 30.35 \\
$k=5$ & PCA    & 58.77 & 87.98 & 88.00 & 87.23 & 79.13 & 81.61 & 70.25 & 56.31 \\ 
      & -      & 65.80 & 85.43 & 86.54 & 85.10 & 81.06 & 80.78 & 38.66 & 32.57 \\
$k=10$& PCA    & 69.66 & 88.76 & 87.53 & 86.17 & 86.64 & 89.07 & 85.34 & 72.47 \\ 
      & -      & 76.26 & 87.92 & 87.47 & 87.29 & 83.71 & 80.43 & 41.50 & 50.65 \\
$k=15$& PCA    & 76.91 & 89.26 & 89.55 & 87.89 & 86.48 & 88.92 & 81.56 & 73.09 \\ 
\hline\hline
\end{tabular}
\end{table}

\begin{table}[t]
\centering
\caption{Purity for the different models using vectors with random sparsification.
Best results are marked in bold for each setting.}
\label{tab2}
\begin{tabular}{llcccccccc}
\hline\hline
      &        & 1     & 2     & 3     & 4     & 5     & 6     & 7     & 8     \\ \hline
$k=5$ & -      & 52.32 & 47.26 & 69.53 & 44.85 & 23.86 & 27.01 & 20.96 & 27.68 \\
      & PCA    & 46.74 & 87.88 & 83.43 & 87.33 & 87.33 & 39.22 & 22.12 & 63.59 \\
      & RK     & 49.28 & 87.35 & 88.91 & 69.03 & 79.05 & 80.85 & 69.06 & 56.58 \\
      & RK-Spat& 49.76 & 87.27 & 87.59 & 86.84 & 79.08 & 73.51 & 68.96 & 56.62 \\[0.2em]
$k=10$& -      & 61.70 & 80.23 & 82.53 & 59.59 & 31.62 & 40.24 & 26.16 & 29.76 \\
      & PCA    & 68.49 & 86.23 & 87.93 & 82.13 & 48.54 & 54.97 & 26.72 & 64.16 \\
      & RK     & 64.81 & 88.50 & 87.13 & 85.16 & 86.23 & 89.13 & 82.47 & 70.96 \\
      & RK-Spat& 68.29 & 88.50 & 86.83 & 87.62 & 86.32 & 89.21 & 83.09 & 68.37 \\[0.2em]
$k=15$& -      & 65.79 & 79.60 & 86.82 & 80.91 & 40.23 & 42.88 & 28.08 & 42.68 \\
      & PCA    & 73.47 & 87.53 & 87.94 & 87.17 & 55.65 & 65.82 & 31.26 & 65.51 \\
      & RK     & 76.58 & 88.95 & 89.45 & 88.97 & 86.96 & 89.26 & 82.34 & 72.64 \\
      & RK-Spat& 71.75 & 88.16 & 90.25 & 86.84 & 85.90 & 89.37 & 82.10 & 72.16 \\
\hline\hline
\end{tabular}
\end{table}

\begin{table}[t]
\centering
\caption{Purity for the different models using vectors with fixed knots sparsification.
Best results are marked in bold for each setting.}
\label{tab3}
\begin{tabular}{llcccccccc}
\hline\hline
      &        & 1     & 2     & 3     & 4     & 5     & 6     & 7     & 8     \\ \hline
$k=5$ & -      & 50.38 & 69.49 & 72.38 & 87.09 & 72.45 & 77.88 & 68.85 & 61.27 \\
      & PCA    & 38.90 & 85.51 & 86.79 & 85.85 & 86.10 & 86.82 & 70.28 & 66.66 \\
      & BS     & 52.24 & 86.13 & 72.54 & 85.83 & 72.79 & 73.38 & 69.42 & 51.29 \\
      & BS-Spat& 49.16 & 86.42 & 86.89 & 85.94 & 61.86 & 86.15 & 69.47 & 51.19 \\[0.2em]
$k=10$& -      & 63.37 & 85.70 & 86.46 & \textbf{84.09} & 84.16 & 82.73 & 76.23 & 67.61 \\
      & PCA    & 59.29 & 81.91 & 85.24 & 81.98 & 84.95 & 87.14 & 82.89 & 71.81 \\
      & BS     & 63.76 & 86.34 & 86.24 & 84.22 & 84.51 & 88.67 & 83.50 & 64.43 \\
      & BS-Spat& 66.44 & 87.71 & 87.53 & 85.12 & 86.06 & 89.29 & 83.36 & 55.64 \\[0.2em]
$k=15$& -      & 71.87 & 87.03 & 86.07 & 86.68 & 83.30 & 86.38 & 63.65 & 72.30 \\
      & PCA    & 65.40 & 88.32 & 85.71 & 84.88 & 85.48 & 87.92 & 81.24 & 75.35 \\
      & BS     & 70.58 & 87.42 & 89.24 & 86.14 & 87.23 & 89.01 & 82.72 & 72.68 \\
      & BS-Spat& 70.67 & 88.24 & 88.61 & 86.73 & 86.47 & 88.54 & 82.76 & 66.99 \\
\hline\hline
\end{tabular}
\end{table}

\section{CONCLUSIONS AND LIMITATION}

In this paper, Random-knots and Random-knots-Spatial estimators are proposed for the covariance of functional data, 
where data is sparsified by randomly intercepting points from each node. The covariance estimator is asymptotically equivalent to
an averaged sample estimator at the rate of $\mathcal{O}_{p}\left(n^{-1}\right)$.
B-spline and Bspline-Spatial estimators are proposed for the covariance of functional data, 
the convergence rate $\mathcal{O}_{p}\left(n^{-1/2}\right)$ of which are derived analogically.
In this case, data is sparsified by intercepting points at fixed positions of each node, and then B-spline interpolation is carried out to avoid the loss of overall data information. We further characterize the uniform weak convergence of the corresponding estimation of eigenvalues and eigenvectors. Spatial factor are necessary to be taken into account when there are
spatial correlations across nodes between the vectors, hence, the standard
approach of simple averaging the sparsified vectors can lead to high estimation error.
Theoretical results are backed by simulation evidence. Main advantage of our
method is its computational efficiency and feasibility for large-scale functional data. It greatly enhances the application of unsupervised domain clustering to multi-domain corpus
in neural machine translation. 

A few more issues still merit further research. For
instance, the knots selection using the AIC works well in practice, but a stronger theoretical
justification for its use is still needed. Our work focuses on the the approximation and theoretical properties of the proposed edtimators, 
while in recent years,
there has been a great deal of work on constructing simultaneous confidence envelopes, which is crucial for making global inference.
It is also interesting to combine our data sparsification methodology with functional regression models and the mean and covariance estimation in such models is a significant challenge
and requires more in‐depth investigation. Last but not
least, extending the novel sparsification procedure for large‐scale
longitudinal data is worth exploring, which is expected to find more applications in various scientific fields. This paper has been submitted as \cite{zhengefficient}.

\bibliography{reference}
\bibliographystyle{plainnat}

\appendix
\section{Appendix}
This section provides technical lemmas and detailed proofs of the main asymptotic results. 
\subsection{Additional Notations}
Throughout this section, $\mathcal{O}_{p}$ (or ${\scriptstyle{\mathcal{O}_{p}}}$ ) denotes a sequence of random variables of certain order in probability. For instance, $\mathcal{O}_{p}\left(n^{-1 / 2}\right)$ means a smaller order than $n^{-1 / 2}$ in probability, and by $\mathcal{O}_{a.s.}$ (or ${\scriptstyle{\mathcal{O}_{a.s.}}}$ ) almost surely $\mathcal{O}$ (or ${\scriptstyle{\mathcal{O}}}$ ). For sequences $a_{n}$ and $b_{n}$, denote $a_{n} \asymp b_{n}$ if $a_{n}$ and $b_{n}$ are asymptotically equivalent. 

For any vector $\mathbf{a}=\left(a_{1}, \ldots, a_{n}\right) \in \mathbb{R}^{n}$, take $\|\mathbf{a}\|_{r}=\left(\left|a_{1}\right|^{r}+\ldots+\left|a_{n}\right|^{r}\right)^{1 / r}$, $1 \leq r<+\infty$, $\|\mathbf{a}\|_{\infty}=\max \left(\left|a_{1}\right|, \ldots,\left|a_{n}\right|\right)$. For any matrix $\mathbf{A}=\left(a_{i j}\right)_{i=1, j=1}^{m, n}$, denote its $L_{r}$ norm as $\|\mathbf{A}\|_{r}=\max _{\mathbf{a} \in R^{n}, \mathbf{a} \neq \mathbf{0}}\|\mathbf{A} \mathbf{a}\|_{r}\|\mathbf{a}\|_{r}^{-1}$, for $r<+\infty$ and $\|\mathbf{A}\|_{r}=\max _{1 \leq i \leq m} \sum_{j=1}^{n}\left|a_{i j}\right|$, for $r=\infty$. For any Lebesgue measurable function $\phi(\mathbf{x})$ on a domain $\mathcal{D}, \mathcal{D}=[0,1]$, let $\|\phi\|_{\infty}=\sup _{\mathbf{x} \in \mathcal{D}}|\phi(\mathbf{x})|$. For any $L^{2}$ integrable functions $\phi(\mathbf{x})$ and $\varphi(\mathbf{x}), \mathbf{x} \in \mathcal{D}$, take $\langle\phi, \varphi\rangle=\int_{\mathcal{D}} \phi(\mathbf{x}) \varphi(\mathbf{x}) d \mathbf{x}$, with $\|\phi\|_{2}^{2}=\langle\phi, \phi\rangle$. We set $\langle\phi, \varphi\rangle_{N}=N^{-1} \sum_{1 \leq j \leq N} \phi\left(\frac{j}{N}\right) \varphi\left(\frac{j}{N}\right)$.

\subsection{Proof of Theorem \ref{Random-k}}
\textsc{Proof.}
The MSE can be computed as
\begin{align*}
\mathbb{E}\|\hat{G}-\bar{G}\|^{2}=&\sum_{j,j^{\prime}=1}^{d}\mathbb{E}\left(\|\frac{1}{n}\left(\frac{d}{J_{s}}\right)^{2}\sum_{i=1}^{n}\left(h_{ij}-\bar{h}_{j}\right)\left(h_{ij^{\prime}}-\bar{h}_{j^{\prime}}\right)-\frac{1}{n}\sum_{i=1}^{n}\left(x_{ij}-\bar{m}_{j}\right)\left(x_{ij^{\prime}}-\bar{m}_{j^{\prime}}\right)\|^{2}\right)
\end{align*}
Now, as $\mathbb{E}\left(\frac{d}{J_{s}} h_{i j}\right)=x_{i j}$ and $\mathbb{E}\left(\frac{d}{J_{s}} h_{i j^{\prime}}\right)=x_{i j^{\prime}}$, it holds that
\begin{align*}
\frac{1}{n^{2}} \sum_{i=1}^{n} &\mathbb{E}\left(\left(\frac{d}{J_{s}}\right)^{2}\left(h_{ij}-\bar{h}_{j}\right)\left(h_{ij^{\prime}}-\bar{h}_{j^{\prime}}\right)-\left(x_{ij}-\bar{m}_{j}\right)\left(x_{ij^{\prime}}-\bar{m}_{j^{\prime}}\right)\right)^{2}\\
=&\frac{1}{n^{2}} \sum_{i=1}^{n}\left(\mathbb{E}\left(\left(\frac{d}{J_{s}}\right)^{2}\left(h_{ij}-\bar{h}_{j}\right)\left(h_{ij^{\prime}}-\bar{h}_{j^{\prime}}\right)\right)^{2}-\mathbb{E}\left(\left(x_{ij}-\bar{m}_{j}\right)\left(x_{ij^{\prime}}-\bar{m}_{j^{\prime}}\right)\right)^{2}\right)
\end{align*}
Since $h_{i j}=x_{i j}$ with probability $J_{s} / d$ and $h_{i j}=0$ otherwise (by definition), therefore 
\begin{align*}
\mathbb{E}\left(\left(\frac{d}{J_{s}}\right)^{2}\left(h_{ij}-\bar{h}_{j}\right)\left(h_{ij^{\prime}}-\bar{h}_{j^{\prime}}\right)\right)^{2}=&\left(\frac{d}{J_{s}}\right)^{4}\mathbb{E}\left(\left(h_{ij}-\bar{h}_{j}\right)\left(h_{ij^{\prime}}-\bar{h}_{j^{\prime}}\right)\right)^{2}\\
=&\left(\frac{d}{J_{s}}\right)^{2}\mathbb{E}\left(\left(x_{ij}-\bar{m}_{j}\right)\left(x_{ij^{\prime}}-\bar{m}_{j^{\prime}}\right)\right)^{2}
\end{align*}
Hence independence of $x_{ij}$ and $x_{ij^{\prime}}$, $1\leq i\leq n$, $j\neq j^{\prime}$ implies
\begin{align*}
\mathbb{E}\|\hat{G}-\bar{G}\|^{2}=&\frac{1}{n^{2}}\left(\left(\frac{d}{J_{s}}\right)^{2}-1\right)\sum_{i=1}^{n}\sum_{j, j^{\prime}=1}^{d}\mathbb{E}\left(\left(x_{ij}-\bar{m}_{j}\right)\left(x_{ij^{\prime}}-\bar{m}_{j^{\prime}}\right)\right)^{2}\\
=&\frac{1}{n^{2}}\left(\left(\frac{d}{J_{s}}\right)^{2}-1\right)\sum_{i=1}^{n}\mathbb{E}\left(\sum_{j=1}^{d}\left(x_{ij}-\bar{m}_{j}\right)^{2}\sum_{j^{\prime}=1}^{d}\left(x_{ij^{\prime}}-\bar{m}_{j^{\prime}}\right)^{2}\right)\\
=&\frac{1}{n^{2}}\left(\left(\frac{d}{J_{s}}\right)^{2}-1\right)R_{1}
\end{align*}
where $R_{1}=\sum_{i=1}^{n}\left\|x_{i}-\bar{m}\right\|^{4}$.

\subsection{Proof of Proposition \ref{THM:Random-k-spatial 1}}
\textsc{Proof.}
Let $\xi_{i j}$ be an indicator random variable which is 1 or 0 , depending on whether $h_{i j}=x_{i j}$ or not for $1\leq i\leq n$, $1\leq j\leq d$.

Case 1: With probability $\left(1-\frac{J_{s}}{d}\right)^{2}$, $\left\{\xi_{i j}=0, \xi_{i j^{\prime}}=0\right\}$ which implies $h_{i j}=0$ and $h_{ij^{\prime}}=0$. Therefore,
\begin{align*}
&\mathbb{E}_{M_{j}, M_{j^{\prime}} \mid \xi_{i j}=0, \xi_{i j^{\prime}}=0}\left[\frac{\bar{\beta}^{2}\left(h_{i j}-\bar{h}_{j}\right)\left(h_{i j^{\prime}}-\bar{h}_{j^{\prime}}\right)}{T\left(M_{j}\right)T\left(M_{j^{\prime}}\right)}\right]\\
=&\mathbb{E}_{M_{j} \mid \xi_{i j}=0}\left[\frac{\bar{\beta}\left(h_{i j}-\bar{h}_{j}\right)}{T\left(M_{j}\right)}\right]\mathbb{E}_{M_{j^{\prime}} \mid \xi_{i j^{\prime}}=0}\left[\frac{\bar{\beta} \left(h_{i j^{\prime}}-\bar{h}_{j^{\prime}}\right)}{T\left(M_{j^{\prime}}\right)}\right]\\
=&0.
\end{align*}

Case 2: With probability $\left(\frac{J_{s}}{d}\right)\left(1-\frac{J_{s}}{d}\right)$, $\left\{\xi_{i j}=0, \xi_{i j^{\prime}}=1\right\}$ which implies $h_{i j}=0$ and $h_{ij^{\prime}}=x_{ij^{\prime}}$. Still we have
$$
\mathbb{E}_{M_{j}, M_{j^{\prime}} \mid \xi_{i j}=0, \xi_{i j^{\prime}}=1}\left[\frac{\bar{\beta}^{2}\left(h_{i j}-\bar{h}_{j}\right)\left(h_{i j^{\prime}}-\bar{h}_{j^{\prime}}\right)}{T\left(M_{j}\right)T\left(M_{j^{\prime}}\right)}\right]=0.
$$

Case 3: With probability $\left(\frac{J_{s}}{d}\right)^{2}$, $\left\{\xi_{i j}=1, \xi_{^{\prime}}=1\right\}$ which implies $h_{i j}=x_{i j}$ and $h_{i j^{\prime}}=x_{i j^{\prime}}$. Therefore,
\begin{align*}
&\mathbb{E}_{M_{j}, M_{j^{\prime}} \mid \xi_{i j}=1,  \xi_{i j^{\prime}}=1}\left[\frac{\bar{\beta}^{2}\left(h_{i j}-\bar{h}_{j}\right)\left(h_{i j^{\prime}}-\bar{h}_{j^{\prime}}\right)}{T\left(M_{j}\right)T\left(M_{j^{\prime}}\right)}\right]\\
=&\mathbb{E}_{M_{j}, M_{j^{\prime}} \mid M_{j} \geq 1, M_{j^{\prime}} \geq 1}\left[\frac{\bar{\beta}^{2}\left(h_{i j}-\bar{h}_{j}\right)\left(h_{i j^{\prime}}-\bar{h}_{j^{\prime}}\right)}{T\left(M_{j}\right)T\left(M_{j^{\prime}}\right)}\right]\\
=&\bar{\beta}^{2} \left(x_{i j}-\bar{m}_{j}\right)\left(x_{i j^{\prime}}-\bar{m}_{j^{\prime}}\right) \mathbb{E}_{M_{j}, M_{j^{\prime}} \mid M_{j} \geq 1, M_{j^{\prime}} \geq 1}\left[\frac{1}{T\left(M_{j}\right)T\left(M_{j^{\prime}}\right)}\right]
\end{align*}
The crucial observation here is that $\xi_{i j}=1$ only implies $M_{j} \geq 1$ and does not give any other information about $M_{j}$. Taking expectation with respect to $\xi_{i j}$ we have,
\begin{align*}
&\mathbb{E}_{\xi_{i j}, \xi_{i j^{\prime}}} \mathbb{E}_{M_{j}, M_{j^{\prime}} \mid \xi_{i j}, \xi_{i j^{\prime}}}\left[\frac{\bar{\beta}^{2}\left(h_{i j}-\bar{h}_{j}\right)\left(h_{i j^{\prime}}-\bar{h}_{j^{\prime}}\right)}{T\left(M_{j}\right)T\left(M_{j^{\prime}}\right)}\right]\\
=&\left(\frac{J_{s}}{d}\right)^{2} \bar{\beta}^{2}\left(x_{i j}-\bar{m}_{j}\right)\left(x_{i j^{\prime}}-\bar{m}_{j^{\prime}}\right) \mathbb{E}_{M_{j}, M_{j^{\prime}} \mid M_{j} \geq 1, M_{j^{\prime}} \geq 1}\left[\frac{1}{T\left(M_{j}\right)T\left(M_{j^{\prime}}\right)}\right]\\
=&\left(x_{i j}-\bar{m}_{j}\right)\left(x_{i j^{\prime}}-\bar{m}_{j^{\prime}}\right)
\end{align*}
which follows from the definition of $\bar{\beta}$ in (\ref{def:betabar}). This proves Proposition \ref{THM:Random-k-spatial 1}.

\subsection{Proof of Theorem \ref{THM:Random-k-spatial 2}}
\textsc{Proof.}
MSE can be computed as
\begin{align}
\label{sub6}
&\mathbb{E}\|\hat{G}-\bar{G}\|^{2}=\sum_{j, j^{\prime}=1}^{d} \mathbb{E}\left(\hat{G}_{jj^{\prime}}-\bar{G}_{jj^{\prime}}\right)^{2}\notag\\
=&\sum_{j, j^{\prime}=1}^{d} \mathbb{E}\left(\frac{1}{n} \frac{\bar{\beta}^{2}}{T\left(M_{j}\right)T\left(M_{j^{\prime}}\right)}\sum_{i=1}^{n}\left(h_{ij}-\bar{h}_{j}\right)\left(h_{ij^{\prime}}-\bar{h}_{j^{\prime}}\right)-\frac{1}{n}\sum_{i=1}^{n}\left(x_{ij}-\bar{m}_{j}\right)\left(x_{ij^{\prime}}-\bar{m}_{j^{\prime}}\right)\right)^{2}
\end{align}
As the estimator is designed to be unbiased, i.e., $\mathbb{E}\left(\frac{1}{n} \frac{\bar{\beta}^{2}}{T\left(M_{j}\right)T\left(M_{j^{\prime}}\right)}\sum_{i=1}^{n}\left(h_{ij}-\bar{h}_{j}\right)\left(h_{ij^{\prime}}-\bar{h}_{j^{\prime}}\right)\right)=\frac{1}{n}\sum_{i=1}^{n}\left(x_{ij}-\bar{m}_{j}\right)\left(x_{ij^{\prime}}-\bar{m}_{j^{\prime}}\right)$, it holds that
\begin{align}
\label{sub4}
&\mathbb{E}\left(\frac{1}{n} \frac{\bar{\beta}^{2}}{T\left(M_{j}\right)T\left(M_{j^{\prime}}\right)}\sum_{i=1}^{n}\left(h_{ij}-\bar{h}_{j}\right)\left(h_{ij^{\prime}}-\bar{h}_{j^{\prime}}\right)-\frac{1}{n}\sum_{i=1}^{n}\left(x_{ij}-\bar{m}_{j}\right)\left(x_{ij^{\prime}}-\bar{m}_{j^{\prime}}\right)\right)^{2}\notag\\
=&\frac{1}{n^{2}} \mathbb{E}\left(\frac{\bar{\beta}^{2}}{T\left(M_{j}\right)T\left(M_{j^{\prime}}\right)}\sum_{i=1}^{n}\left(h_{ij}-\bar{h}_{j}\right)\left(h_{ij^{\prime}}-\bar{h}_{j^{\prime}}\right)\right)^{2}-\frac{1}{n^{2}}\left(\sum_{i=1}^{n}\left(x_{ij}-\bar{m}_{j}\right)\left(x_{ij^{\prime}}-\bar{m}_{j^{\prime}}\right)\right)^{2}
\end{align}
We now analyze the first term above.
\begin{align}
\label{sub}
&\mathbb{E}\left(\frac{\bar{\beta}^{2}}{T\left(M_{j}\right)T\left(M_{j^{\prime}}\right)}\sum_{i=1}^{n}\left(h_{ij}-\bar{h}_{j}\right)\left(h_{ij^{\prime}}-\bar{h}_{j^{\prime}}\right)\right)^{2}\notag\\
=&\sum_{i=1}^{n} \bar{\beta}^{4} \mathbb{E}\left[\frac{\left(h_{i j}-\bar{h}_{j}\right)^{2}\left(h_{i j^{\prime}}-\bar{h}_{j^{\prime}}\right)^{2}}{T\left(M_{j}\right)^{2}T\left(M_{j^{\prime}}\right)^{2}}\right]\notag\\
&+2\sum_{i=1}^{n} \sum_{k=i+1}^{n} \bar{\beta}^{4} \mathbb{E}\left[\frac{\left(h_{i j}-\bar{h}_{j}\right)\left( h_{k j}-\bar{h}_{j}\right)\left(h_{i j^{\prime}}-\bar{h}_{j^{\prime}}\right)\left( h_{k j^{\prime}}-\bar{h}_{j^{\prime}}\right)}{T\left(M_{j}\right)^{2}T\left(M_{j^{\prime}}\right)^{2}}\right]
\end{align}
Note here that the expectation is taken over the randomness in $h_{i j}$ as well as $T\left(M_{j}\right)$. Further, $\bar{\beta}^{4}\frac{\left(h_{i j}-\bar{h}_{j}\right)^{2}\left(h_{i j^{\prime}}-\bar{h}_{j^{\prime}}\right)^{2}}{T\left(M_{j}\right)^{2}T\left(M_{j^{\prime}}\right)^{2}}$ is non-zero only when a node $i$ samples coordinate $j$ and $j^{\prime}$, i.e., $h_{i j}=x_{i j}$ and $h_{i j^{\prime}}=x_{i j^{\prime}}$. This implies that $M_{j} \geq 1$ and $M_{j^{\prime}} \geq 1$.
$$
\mathbb{P}\left(M_{j}=m\right)=\binom{n}{m} p^{m}(1-p)^{n-m}$$
where $p=\frac{J_{s}}{d}$.
Therefore, by the law of total expectation, we have
\begin{align}
\label{sub1}
&\bar{\beta}^{4} \mathbb{E}\left[\frac{\left(h_{i j}-\bar{h}_{j}\right)^{2}\left(h_{i j^{\prime}}-\bar{h}_{j^{\prime}}\right)^{2}}{T\left(M_{j}\right)^{2}T\left(M_{j^{\prime}}\right)^{2}}\right]\notag\\ =&\bar{\beta}^{4} \mathbb{E}_{M_{j} \mid M_{j} \geq 1}\left[\frac{J_{s} \left(x_{i j}-\bar{m}_{j}\right)^{2}}{d T\left(M_{j}\right)^{2}}\right]\mathbb{E}_{M_{j^{\prime}} \mid M_{j^{\prime}} \geq 1}\left[\frac{J_{s} \left(x_{i j^{\prime}}-\bar{m}_{j^{\prime}}\right)^{2}}{d T\left(M_{j^{\prime}}\right)^{2}}\right] \notag\\
=&\left(\bar{\beta}^{4}\sum_{r, r^{\prime}=1}^{n} \frac{J_{s}}{d T(r)^{2}}\frac{J_{s}}{d T(r^{\prime})^{2}}\binom{n-1}{r-1}\binom{n-1}{r^{\prime}-1}\left(\frac{J_{s}}{d}\right)^{r+r^{\prime}-2}\left(1-\frac{J_{s}}{d}\right)^{2n-r-r^{\prime}}\right)\notag \\
&\quad\times\left(x_{i j}-\bar{m}_{j}\right)^{2}\left(x_{i j^{\prime}}-\bar{m}_{j^{\prime}}\right)^{2} \notag\\
=&\left(\frac{d}{J_{s}}+c_{1}\right)^{2} \left(x_{i j}-\bar{m}_{j}\right)^{2}\left(x_{i j^{\prime}}-\bar{m}_{j^{\prime}}\right)^{2} 
\end{align}
where $c_{1}=\bar{\beta}^{2} \sum_{r=1}^{n} \frac{J_{s}}{d T(r)^{2}}\binom{n-1}{r-1}\left(\frac{J_{s}}{d}\right)^{r-1}\left(1-\frac{J_{s}}{d}\right)^{n-r}-\frac{d}{J_{s}}$. Here, the second equality uses the fact that when node $i$ samples coordinate $j$ and $j^{\prime}$ (i.e., $x_{i j}=h_{i j}$, $x_{i j^{\prime}}=h_{i j^{\prime}}$ ), then $M_{j} \geq 1$ and $M_{j^{\prime}} \geq 1$.

Following a similar argument as above, note that $\frac{\left(h_{i j}-\bar{h}_{j}\right)\left( h_{k j}-\bar{h}_{j}\right)\left(h_{i j^{\prime}}-\bar{h}_{j^{\prime}}\right)\left( h_{k j^{\prime}}-\bar{h}_{j^{\prime}}\right)}{T\left(M_{j}\right)^{2}T\left(M_{j^{\prime}}\right)^{2}}$ is non-zero only when nodes $i$ and $J_{s}$ sample coordinate $j$ and $j^{\prime}$, i.e., $h_{i j}=x_{i j}$, $h_{k j}=x_{k j}$, $h_{i j^{\prime}}=x_{i j^{\prime}}$, $h_{k j^{\prime}}=x_{k j^{\prime}}$. This implies that $M_{j} \geq 2$ and $M_{j^{\prime}} \geq 2$. Therefore, by the law of total expectation, we have
\begin{align}
\label{sub2}
&\bar{\beta}^{4}\mathbb{E}\left(\frac{\left(h_{i j}-\bar{h}_{j}\right)\left( h_{k j}-\bar{h}_{j}\right)\left(h_{i j^{\prime}}-\bar{h}_{j^{\prime}}\right)\left( h_{k j^{\prime}}-\bar{h}_{j^{\prime}}\right)}{T\left(M_{j}\right)^{2}T\left(M_{j^{\prime}}\right)^{2}}\right)\notag\\
=&\bar{\beta}^{4}\mathbb{E}_{M_{j} \mid M_{j} \geq 2}\left(\frac{J_{s} \left(x_{i j}-\bar{m}_{j} \right)\left(x_{k j}-\bar{m}_{j}\right)}{d T\left(M_{j}\right)^{2}}\right)\mathbb{E}_{M_{j^{\prime}} \mid M_{j^{\prime}} \geq 2}\left(\frac{J_{s} \left(x_{i j^{\prime}}-\bar{m}_{j^{\prime}} \right)\left(x_{k j^{\prime}}-\bar{m}_{j^{\prime}}\right)}{d T\left(M_{j^{\prime}}\right)^{2}}\right) \notag\\
=&\left(\bar{\beta}^{4}\sum_{r, r^{\prime}=2}^{n} \frac{J_{s}}{d T\left(r^{\prime}\right)^{2}}\frac{J_{s}^{2}}{d^{2} T\left(r^{\prime}\right)^{2}}\binom{n-2}{r-2}\binom{n-2}{r^{\prime}-2}\left(\frac{J_{s}^{2}}{d^{2}}\right)^{r+r^{\prime}-4}\left(1-\frac{J_{s}}{d}\right)^{2n-r-r^{\prime}}\right)\notag\\
&\quad\times\left(x_{i j}-\bar{m}_{j}\right) \left(x_{k j}-\bar{m}_{j}\right) \left(x_{i j^{\prime}}-\bar{m}_{j^{\prime}}\right) \left(x_{k j^{\prime}}-\bar{m}_{j^{\prime}}\right) \notag\\
=&\left(1-c_{2}\right)^{2}\left(x_{i j}-\bar{m}_{j}\right) \left(x_{k j}-\bar{m}_{j}\right) \left(x_{i j^{\prime}}-\bar{m}_{j^{\prime}}\right) \left(x_{k j^{\prime}}-\bar{m}_{j^{\prime}}\right)
\end{align}
where $c_{2}=1-\bar{\beta}^{2} \sum_{r=2}^{n} \frac{J_{s}^{2}}{d^{2} T\left(r\right)^{2}}\binom{n-2}{r-2}\left(\frac{J_{s}}{d}\right)^{r-2}\left(1-\frac{J_{s}}{d}\right)^{n-r}$.

Substituting (\ref{sub1}) and (\ref{sub2}) in (\ref{sub}), we get
\begin{align}
\label{sub3}
&\mathbb{E}\left(\frac{\bar{\beta}^{2}}{T\left(M_{j}\right)T\left(M_{j^{\prime}}\right)}\sum_{i=1}^{n}\left(h_{ij}-\bar{h}_{j}\right)\left(h_{ij^{\prime}}-\bar{h}_{j^{\prime}}\right)\right)^{2}\notag\\
=&\left(\frac{d}{J_{s}}+c_{1}\right)^{2} \sum_{i=1}^{n}\left(x_{i j}-\bar{m}_{j}\right)^{2}\left(x_{i j^{\prime}}-\bar{m}_{j^{\prime}}\right)^{2}\notag\\
&+\left(1-c_{2}\right)^{2}\sum_{i=1}^{n} \sum_{k=i+1}^{n}\left(x_{i j}-\bar{m}_{j}\right) \left(x_{k j}-\bar{m}_{j}\right) \left(x_{i j^{\prime}}-\bar{m}_{j^{\prime}}\right) \left(x_{k j^{\prime}}-\bar{m}_{j^{\prime}}\right)
\end{align}
Now, substituting (\ref{sub3}) in (\ref{sub4}), we get
\begin{align}
\label{sub5}
&\mathbb{E}\left(\frac{1}{n} \frac{\bar{\beta}^{2}}{T\left(M_{j}\right)T\left(M_{j^{\prime}}\right)}\sum_{i=1}^{n}\left(h_{ij}-\bar{h}_{j}\right)\left(h_{ij^{\prime}}-\bar{h}_{j^{\prime}}\right)-\frac{1}{n}\sum_{i=1}^{n}\left(x_{ij}-\bar{m}_{j}\right)\left(x_{ij^{\prime}}-\bar{m}_{j^{\prime}}\right)\right)^{2}\notag\\
=&\frac{1}{n^{2}}\left(\left(\frac{d}{J_{s}}+c_{1}\right)^{2}-1\right) \sum_{i=1}^{n} \left(x_{i j}-\bar{m}_{j}\right)^{2}\left(x_{i j^{\prime}}-\bar{m}_{j^{\prime}}\right)^{2}\notag\\
&+\frac{1}{n^{2}} \left(\left(1-c_{2}\right)^{2}-1\right) \sum_{i=1}^{n} \sum_{k=i+1}^{n} \left(x_{i j}-\bar{m}_{j}\right) \left(x_{k j}-\bar{m}_{j}\right)\left(x_{i j^{\prime}}-\bar{m}_{j^{\prime}}\right) \left(x_{k j^{\prime}}-\bar{m}_{j^{\prime}}\right)
\end{align}
Finally replacing (\ref{sub5}) in (\ref{sub6}) we get,
\begin{align*}
\mathbb{E}\|\hat{G}-\bar{G}\|^{2}=\frac{1}{n^{2}}\left(\left(\frac{d}{J_{s}}+c_{1}\right)^{2}-1\right) R_{1}+\frac{1}{n^{2}}\left(\left(1-c_{2}\right)^{2}-1\right) R_{2}
\end{align*}
where $R_{1}=\sum_{i=1}^{n}\left\|x_{i}-\bar{m}\right\|^{4}$ and $R_{2}=2 \sum_{i}^{n} \sum_{k=i+1}^{n}\left\langle\left(x_{i}-\bar{m}\right)^{2}, \left(x_{k}-\bar{m}\right)^{2}\right\rangle$.

\subsection{Proof of Theorem \ref{THM:Random-k-spatial 3}}
\textsc{Proof.}
Observe that in (\ref{diff}), the only term that depends on $T(\cdot)$ is 
$$\left(c_{1}^{2}+2c_{1}\frac{d}{J_{s}}\right)R_{1}+\left(c_{2}^{2}-2c_{2}\right)R_{2}=\left(c_{1}+\frac{d}{J_{s}}\right)^{2}R_{1}+\left(c_{2}-1\right)^{2}R_{2}-\left(\frac{d}{J_{s}}\right)^{2}R_{1}-R_{2}$$
Thus to find the function $T^{*}(\cdot)$ that minimizes the MSE, we just need to minimize this term.

Next, from the definitions of $c_{1}$ and $c_{2}$ in Theorem \ref{THM:Random-k-spatial 2}, we can obtain the following expression for $T^{*}(\cdot)$
\begin{align}
\label{optT}
T^{*}(r) &=\underset{T}{\arg \min } \bar{\beta}^{4} \left(\sum_{r=1}^{n} \frac{J_{s}}{d T(r)^{2}}\binom{n-1}{r-1}\left(\frac{J_{s}}{d}\right)^{r-1}\left(1-\frac{J_{s}}{d}\right)^{n-r}\right)^{2} \notag\\
&+\frac{R_{2}}{R_{1}} \bar{\beta}^{4} \left(\sum_{r=2}^{n} \frac{J_{s}^{2}}{d^{2} T(r)^{2}}\binom{n-2}{r-2}\left(\frac{J_{s}}{d}\right)^{r-2}\left(1-\frac{J_{s}}{d}\right)^{n-r}\right)^{2}-\frac{R_{2}}{R_{1}}.
\end{align}

We claim that $T^{*}\left(r\right)=\left(1+\frac{R_{2}}{R_{1}} \left(\frac{r-1}{n-1}\right)^{2}\right)^{1/2}$ is an optimal solution for our objective defined in (\ref{optT}). To see this, consider the following cases,

Case 1: $p=0$ or $p=1$.
In this case $c_{1}$ and $c_{2}$ are independent of $T(\cdot)$ and hence our objective does not depend on the choice of $T(\cdot)$.

Case 2: $0<p<1$, we define
\begin{align}
\label{w}
\mathbf{w}^{*}=\underset{\mathbf{w}}{\arg \min } \frac{\mathbf{w}^{\top} \mathbf{A} \mathbf{w}}{\left(\mathbf{b}^{\top} \mathbf{w}\right)^{2}},
\end{align}
where $\mathbf{w}$ is a $n$-dimensional vector whose $r$-th entry is $w_{r}=1/T(r)^{2}$, $\mathbf{b}$ is a vector whose $r$-th entry is
$$
b_{r}=\left(\binom{n-1}{r-1} p^{r-1}(1-p)^{n-r}\right)^{2}
$$
where $p=J_{s} / d$, and $\mathbf{A}$ is a diagonal matrix whose $r$-th diagonal entry is
\begin{align*}
A_{r r} &=\left(\binom{n-1}{r-1} p^{r-1}(1-p)^{n-r}\right)^{2}+\frac{R_{2}}{R_{1}}\left(p \binom{n-2}{r-2}p^{r-2}(1-p)^{n-r}\right)^{2} \\
&=b_{r}\left(1+\frac{R_{2}}{R_{1}}\left( \frac{r-1}{n-1}\right)^{2}\right).
\end{align*}
Note that $A_{r r}>0$ for all $r \in\{1, \ldots, n\}$ which implies that $\mathbf{w} \rightarrow \mathbf{A}^{1 / 2} \mathbf{w}$ is a one-to-one mapping. Therefore setting $\mathbf{z}=\mathbf{A}^{1 / 2} \mathbf{w}$, the objective in (\ref{w}) reduces to
\begin{align}
\label{z}
\mathbf{z}^{*}=\underset{\mathbf{z}}{\arg \min } \frac{\|\mathbf{z}\|^{2}}{\left(\mathbf{b}^{\top} \mathbf{A}^{-1 / 2} \mathbf{z}\right)^{2}}
\end{align}
Observe that the objectives (\ref{w}), (\ref{z}) are invariant to the scale of $T(\cdot)$, $\mathbf{w}$, and $\mathbf{z}$ respectively and thus the solutions will be unique up to a scaling factor. Therefore, in the case of (\ref{z}), it is sufficient to solve the reduced objective,
$$
\mathbf{z}^{*}=\underset{\mathbf{z},\|\mathbf{z}\|=1}{\arg \min } \frac{\|\mathbf{z}\|^{2}}{\left(\mathbf{b}^{\top} \mathbf{A}^{-1 / 2} \mathbf{z}\right)^{2}}=\underset{\mathbf{z},\|\mathbf{z}\|=1}{\arg \min } \frac{1}{\left(\mathbf{b}^{\top} \mathbf{A}^{-1 / 2} \mathbf{z}\right)^{2}}
$$
which is minimized (denominator is maximized) by $\mathbf{z}^{*}=\frac{\mathbf{A}^{-1 / 2} \mathbf{b}}{\left\|\mathbf{A}^{-1 / 2} \mathbf{b}\right\|}$. Therefore, the optimal solution (up to a constant) is $\mathbf{w}^{*}=\mathbf{A}^{-1 / 2}\left(\mathbf{A}^{-1 / 2} \mathbf{b}\right)$. Correspondingly, we have that
$$
T^{*}(r)=\left(w_{r}^{*}\right)^{-1/2}=\left(\frac{A_{r r}}{b_{r}}\right)^{1/2}=\left(1+\frac{R_{2}}{R_{1}}\left( \frac{r-1}{n-1}\right)^{2}\right)^{1/2} .
$$
minimizes (\ref{optT}), and consequently minimizes the MSE of the Random-knots-Spatial estimator.

\subsection{Proof of Theorem \ref{THM:spline}}
\begin{lemma}
Let $W_{i} \sim N\left(0, \sigma_{i}^{2}\right), \sigma_{i}>0, i=1, \ldots, n$, then for $n>2, a>2$
$$
\mathbb{P}\left(\max _{1 \leq i \leq n}\left|W_{i} / \sigma_{i}\right|>a \sqrt{\log n}\right)<\sqrt{\frac{2}{\pi}} n^{1-a^{2} / 2} .
$$
Hence, $\left(\max _{1 \leq i \leq n}\left|W_{i}\right|\right) /\left(\max _{1 \leq i \leq n} \sigma_{i}\right) \leq \max _{1 \leq i \leq n}\left|W_{i} / \sigma_{i}\right|=\mathcal{O}_{a . s .}(\sqrt{\log n})$.
\end{lemma}
\textsc{Proof.}
Note that
$$
\begin{aligned}
\mathbb{P}\left(\max _{1 \leq i \leq n}\left|\frac{W_{i}}{\sigma_{i}}\right|>a \sqrt{\log n}\right) & \leq \sum_{i=1}^{n} \mathbb{P}\left(\left|\frac{W_{i}}{\sigma_{i}}\right|>a \sqrt{\log n}\right) \\
& \leq 2 n\{1-\Phi(a \sqrt{\log n})\}<2 n \frac{\phi(a \sqrt{\log n})}{a \sqrt{\log n}} \\
& \leq 2 n \phi(a \sqrt{\log n})=\sqrt{\frac{2}{\pi}} n^{1-a^{2} / 2}
\end{aligned}
$$
for $n>2, a>2$. The lemma follows by applying Borel-Cantelli Lemma.

\begin{lemma}
\label{xhat}
As $n \rightarrow \infty$, we have
$$
\max _{1 \leq i \leq n}\left\|h_{i}-x_{i}\right\|_{\infty}=\mathcal{O}_{a . s .}\left\{J_{s}^{-p^{*}}(n \log n)^{2 / r_{0}}\right\}={\scriptstyle{\mathcal{O}}}_{a.s.}\left(n^{-1 / 2}\right) .
$$
\end{lemma}
\textsc{Proof.}
The trajectory $x_{i}(t)$ is written as $x_{i}(t)=m(t)+\sum_{k=1}^{\infty} \xi_{i k} \phi_{k}(t)$. Denote $\boldsymbol{\phi}_{k}=\left(\phi_{k}(1 / d), \ldots, \phi_{k}(d / d)\right)^{\top}$, and let $\hat{\phi}_{k}(t)=d^{-1} \mathbf{B}(t)^{\top} \mathbf{V}_{n, p}^{-1} \mathbf{B}^{\top} \phi_{k}$ be the B-spline smoothing of $\phi_{k}(t)$. The linearity of spline smoothing implies that
$$
h_{i}(t)-x_{i}(t)=\hat{m}(t)-m(t)+\sum_{k=1}^{\infty} \xi_{i k}\left\{\hat{\phi}_{k}(t)-\phi_{k}(t)\right\} .
$$
Lemma A.4 in \cite{Cao12} assures there exists a constant $C_{q, \mu}>0$, such that
\begin{align}
&\|\hat{m}-m\|_{\infty} \leq C_{q, \mu}\|m\|_{q, \mu} J_{s}^{-p^{*}}, \label{mhat}\\
&\left\|\hat{\phi}_{k}-\phi_{k}\right\|_{\infty} \leq C_{q, \mu}\left\|\phi_{k}\right\|_{q, \mu} J_{s}^{-p^{*}}, \quad k \geq 1 \label{phihat}
\end{align}
Thus, with norm inequality, we have
$$
\left\|h_{i}-x_{i}\right\|_{\infty} \leq\|\hat{m}-m\|_{\infty}+\sum_{k=1}^{\infty}\left|\xi_{i k}\right|\left\|\hat{\phi}_{k}-\phi_{k}\right\|_{\infty} \leq C_{q, \mu} W_{i} J_{s}^{-p^{*}}
$$
where $W_{i}=\|m\|_{q, \mu}+\sum_{k=1}^{\infty}\left|\xi_{i k}\right|\left\|\phi_{k}\right\|_{q, \mu}, i=1, \ldots, n$, are i.i.d. nonnegative random variables. $W_{i}^{r_{0}}$ has a finite absolute moment and we have
$$
\mathbb{P}\left\{\max _{1 \leq i \leq n} W_{i}>(n \log n)^{2 / r_{0}}\right\} \leq n \frac{\mathbb{E} W_{i}^{r_{0}}}{(n \log n)^{2}}=\mathbb{E} W_{i}^{r_{0}}(n \log n)^{-2}
$$
which implies
$$
\sum_{n=1}^{\infty} \mathbb{P}\left\{\max _{1 \leq i \leq n} W_{i}>(n \log n)^{2 / r_{0}}\right\} \leq \mathbb{E} W_{i}^{r_{0}} \sum_{n=1}^{\infty}(n \log n)^{-2}<+\infty
$$
According to Borel Cantelli lemma, $\max _{1 \leq i \leq n} W_{i}=\mathcal{O}_{a . s .}\left\{(n \log n)^{2 / r_{0}}\right\}$ which, together with (\ref{mhat}) and (\ref{phihat}), prove the Lemma \ref{xhat}.

Consequently, the approximation error of $\hat{m}(\cdot)-\bar{m}(\cdot)$ can be decomposed as
\begin{align*}
\hat{m}(\cdot)-\bar{m}(\cdot) &=n^{-1} \sum_{i=1}^{n}\left\{h_{i}(\cdot)-x_{i}(\cdot)\right\}
\end{align*}
According to Lemma \ref{xhat},
\begin{align*}
\sup _{t \in[0,1]} n^{1 / 2}|\hat{m}(t)-\bar{m}(t)| & \leq n^{1 / 2} \max _{1 \leq i \leq n}\left\|h_{i}-x_{i}\right\|_{\infty} =o_{\text {a.s. }}(1).
\end{align*}
(\ref{THM:splinemean}) is proved.

\begin{lemma}
\label{Zhat}
As $n \rightarrow \infty$
\begin{align*}
&\max _{1 \leq i \leq n}\left\|\hat{Z}_{i}-Z_{i}\right\|_{\infty}=\mathcal{O}_{\text {a.s. }}\left\{J_{s}^{-p^{*}}(n \log n)^{2 / r_{0}}\right\}, \\
&\max _{1 \leq i \leq n}\left\|Z_{i}\right\|_{\infty}=\mathcal{O}_{\text {a.s. }}\left\{(n \log n)^{2 / r_{0}}\right\}
\end{align*}
\end{lemma}

\textsc{Proof.} 
Denote $\hat{\phi}_{k}(x)=d^{-1} \mathbf{B}(x)^{\top} \mathbf{V}_{n, p}^{-1} \mathbf{B}^{\top} \phi_{k}$ and $\hat{Z}_{i}(t)=\sum_{k=1}^{\infty} \xi_{i k} \hat{\phi}_{k}(t)$ for $k \in \mathbb{N}_{+}$, hence,
$$
\hat{Z}_{i}(t)-Z_{i}(t)=\sum_{k=1}^{\infty} \xi_{i k}\left\{\hat{\phi}_{k}(t)-\phi_{k}(t)\right\} .
$$
By (\ref{phihat}),
$$
\left\|\hat{Z}_{i}-Z_{i}\right\|_{\infty} \leq \sum_{k=1}^{\infty}\left|\xi_{i k}\right|\left\|\hat{\phi}_{k}-\phi_{k}\right\|_{\infty} \leq C W_{i} J_{s}^{-p^{*}} \text {, }
$$
where $W_{i}=\sum_{k=1}^{\infty}\left|\xi_{i k}\right|\left\|\phi_{k}\right\|_{q, \mu}, i=1, \ldots, n$, are i.i.d nonnegative random variables with finite absolute moment. Then
$$
\mathbb{P}\left\{\max _{1 \leq i \leq n} W_{i}>(n \log n)^{2 / r_{0}}\right\} \leq n \frac{\mathbb{E} W_{i}^{r_{0}}}{(n \log n)^{2}}=\mathbb{E} W_{i}^{r_{0}} n^{-1}(\log n)^{-2},
$$
thus,
$$
\sum_{n=1}^{\infty} \mathbb{P}\left\{\max _{1 \leq i \leq n} W_{i}>(n \log n)^{2 / r_{0}}\right\} \leq \mathbb{E} W_{i}^{r_{0}} \sum_{n=1}^{\infty} n^{-1}(\log n)^{-2}<+\infty,
$$
so $\max _{1 \leq i \leq n} W_{i}=\mathcal{O}_{\text {a.s. }}\left\{(n \log n)^{2 / r_{0}}\right\}$. Similarly, one obtains $\max _{1 \leq i \leq n}\left\|Z_{i}\right\|_{\infty}=\mathcal{O}_{\text {a.s. }}\left\{(n \log n)^{2 / r_{0}}\right\}$. Lemma \ref{Zhat} is obtained.

For any $t, t^{\prime} \in[0,1]$, one could decompose $\hat{G}\left(t, t^{\prime}\right)-\bar{G}\left(t, t^{\prime}\right)$ into three parts
\begin{align*}
\hat{G}\left(t, t^{\prime}\right)-\bar{G}\left(t, t^{\prime}\right) &=n^{-1} \sum_{i=1}^{n} \hat{Z}_{i}(t) \hat{Z}_{i}\left(t^{\prime}\right)-n^{-1} \sum_{i=1}^{n} \bar{Z}_{i}(t) \bar{Z}_{i}\left(t^{\prime}\right) \\
&=\mathrm{I}\left(t, t^{\prime}\right)+\operatorname{II}\left(t, t^{\prime}\right)+\operatorname{III}\left(t, t^{\prime}\right)
\end{align*}
where
\begin{align*}
&\operatorname{I}\left(t, t^{\prime}\right)=n^{-1} \sum_{i=1}^{n}\left\{\hat{Z}_{i}(t)-\bar{Z}_{i}(t)\right\}\left\{\hat{Z}_{i}\left(t^{\prime}\right)-\bar{Z}_{i}\left(t^{\prime}\right)\right\} \\
&\operatorname{II}\left(t, t^{\prime}\right)=n^{-1} \sum_{i=1}^{n} \bar{Z}_{i}\left(t^{\prime}\right)\left\{\hat{Z}_{i}(t)-\bar{Z}_{i}(t)\right\} \\
&\operatorname{III}\left(t, t^{\prime}\right)=n^{-1} \sum_{i=1}^{n} \bar{Z}_{i}(t)\left\{\hat{Z}_{i}\left(t^{\prime}\right)-\bar{Z}_{i}\left(t^{\prime}\right)\right\} .
\end{align*}
According to decomposition of $\left\{h_{i}\right\}_{i=1}^{n}$ and $\left\{x_{i}\right\}_{i=1}^{n}$, one obtains $h_{i}(t)-x_{i}(t)=\hat{Z}_{i}(t)-Z_{i}(t)+\hat{m}(t)-m(t)$, then $\hat{Z}_{i}(t)-\bar{Z}_{i}(t)$ can be represented by
\begin{align*}
\hat{Z}_{i}(t)-\bar{Z}_{i}(t) &=h_{i}(t)-\hat{m}(t)-\left\{x_{i}(t)-\bar{m}(t)\right\} \\
&=h_{i}(t)-n^{-1} \sum_{i^{\prime}=1}^{n} h_{i^{\prime}}(t)-\left\{x_{i}(t)-n^{-1} \sum_{i^{\prime}=1}^{n} x_{i^{\prime}}(t)\right\} \\
&=h_{i}(t)-x_{i}(t)-n^{-1} \sum_{i^{\prime}=1}^{n}\left\{h_{i^{\prime}}(t)-x_{i^{\prime}}(t)\right\} \\
&=\hat{Z}_{i}(t)-Z_{i}(x)-n^{-1} \sum_{i^{\prime}=1}^{n}\left\{\hat{Z}_{i^{\prime}}(t)-Z_{i^{\prime}}(t)\right\} \\
&=\hat{Z}_{i}(t)-Z_{i}(t)-\Theta_{1}(t) .
\end{align*}
and then
\begin{align*}
\mathrm{I}\left(t, t^{\prime}\right)=& n^{-1} \sum_{i=1}^{n}\left\{\hat{Z}_{i}(t)-Z_{i}(t)-\Theta_{1}(t)\right\}\left\{\hat{Z}_{i}\left(t^{\prime}\right)-Z_{i}\left(t^{\prime}\right)-\Theta_{1}\left(t^{\prime}\right)\right\} \\
=& n^{-1} \sum_{i=1}^{n}\left\{\hat{Z}_{i}(t)-Z_{i}(t)\right\}\left\{\hat{Z}_{i}\left(t^{\prime}\right)-Z_{i}\left(t^{\prime}\right)\right\} -n^{-1} \sum_{i=1}^{n}\left\{\hat{Z}_{i}(t)-Z_{i}(t)\right\} \Theta_{1}\left(t^{\prime}\right) \\
&-n^{-1} \sum_{i=1}^{n} \Theta_{1}(t)\left\{\hat{Z}_{i}\left(t^{\prime}\right)-Z_{i}\left(t^{\prime}\right)\right\}+\Theta_{1}(t) \Theta_{1}\left(t^{\prime}\right) \\
=& \Theta_{2}\left(t, t^{\prime}\right)-\Theta_{1}(t) \Theta_{1}\left(t^{\prime}\right) .
\end{align*}
where
$$
\begin{aligned}
&\Theta_{1}(t)=n^{-1} \sum_{i=1}^{n}\left\{\hat{Z}_{i}(t)-Z_{i}(t)\right\} \\
&\Theta_{2}\left(t, t^{\prime}\right)=n^{-1} \sum_{i=1}^{n}\left\{\hat{Z}_{i}(t)-Z_{i}(t)\right\}\left\{\hat{Z}_{i}\left(t^{\prime}\right)-Z_{i}\left(t^{\prime}\right)\right\} .
\end{aligned}
$$
According to Lemma \ref{Zhat} and Assumption 6,
\begin{align*}
&\Theta_{1}(t) \leq \max _{1 \leq i \leq n}\left\|\hat{Z}_{i}-Z_{i}\right\|_{\infty}=\mathcal{O}_{\text {a.s. }}\left\{J_{s}^{-p^{*}}(n \log n)^{2 / r_{0}}\right\}=\mathcal{O}_{\text {a.s. }}\left(n^{-1 / 2}\right)\\
&\Theta_{2}\left(t, t^{\prime}\right)\leq \left(\max _{1 \leq i \leq n}\left\|\hat{Z}_{i}-Z_{i}\right\|_{\infty}\right)^{2}=\mathcal{O}_{\text {a.s. }}\left\{J_{s}^{-2 p^{*}}(n \log n)^{4 / r_{0}}\right\}=\mathcal{O}_{\text {a.s. }}\left(n^{-1 / 2}\right)
\end{align*}
Hence, one obtains $\sup _{t, t^{\prime} \in[0,1]}\left|\mathrm{I}\left(t, t^{\prime}\right)\right|=\mathcal{O}_{a . s .}\left(n^{-1 / 2}\right)$. Moreover,
$$
\begin{aligned}
\mathrm{II}\left(t, t^{\prime}\right)=& n^{-1} \sum_{i=1}^{n}\left\{Z_{i}\left(t^{\prime}\right)-n^{-1} \sum_{i=1}^{n} Z_{i}\left(t^{\prime}\right)\right\}\left\{\hat{Z}_{i}(t)-Z_{i}(t)-\Theta_{1}(t)\right\} \\
=& n^{-1}\sum_{i=1}^{n} Z_{i}\left(x^{\prime}\right)\left\{\hat{Z}_{i}(x)-Z_{i}(x)\right\}-n^{-2}\left[\sum_{i=1}^{n} Z_{i}\left(x^{\prime}\right) \sum_{i^{\prime}=1}^{n} \hat{Z}_{i}(x)+\sum_{i=1}^{n} Z_{i}\left(x^{\prime}\right) \sum_{i^{\prime}=1}^{n} Z_{i}\left(x^{\prime}\right)\right] .
\end{aligned}
$$
By Lemma \ref{Zhat}, one obtains $\sup _{t, t^{\prime} \in[0,1]}\left|\mathrm{II}\left(t, t^{\prime}\right)\right|=\mathcal{O}_{\text {a.s. }}\left(n^{-1 / 2}\right)$. Similarly, $\sup _{t, t^{\prime} \in[0,1]}\left|\operatorname{III}\left(t, t^{\prime}\right)\right|=\mathcal{O}_{\text {a.s. }}\left(n^{-1 / 2}\right)$. Consequently,
$$
\sup _{t, t^{\prime} \in[0,1]}\left|\hat{G}\left(t, t^{\prime}\right)-\bar{G}\left(t, t^{\prime}\right)\right|=\mathcal{O}_{a . s .}\left(n^{-1 / 2}\right) .
$$

\subsection{Proof of Theorem \ref{THM:spline-spatial}}

\textsc{Proof.}
The estimation error of spatial mean can be computed as
\begin{align*}
\|\hat{m}-\bar{m}\|_{\infty}=&\max_{1\leq j \leq d} \left\|\frac{1}{n} \frac{\bar{\beta}}{T\left(M_{j}\right)} \sum_{i=1}^{n} h_{i j}-\frac{1}{n} \sum_{i=1}^{n} x_{i j}\right\|\\
=&\frac{1}{n}\max_{1\leq j\leq d} \left\| \frac{\bar{\beta}}{T\left(M_{j}\right)} \sum_{i=1}^{n} h_{i j}-\frac{\bar{\beta}}{T\left(M_{j}\right)} \sum_{i=1}^{n} x_{i j}+\frac{\bar{\beta}}{T\left(M_{j}\right)} \sum_{i=1}^{n} x_{i j}- \sum_{i=1}^{n} x_{i j}\right\|\\
\leq &\frac{\bar{\beta}}{T\left(M_{j}\right)}\left\|\hat{m}-\bar{m}\right\|_{\infty}+\left(\frac{\bar{\beta}}{T\left(M_{j}\right)}-1\right)\left\|\bar{m}\right\|_{\infty}\\
=& {\scriptstyle{\mathcal{O}}}_{p}\left(n^{-1 / 2}\right) 
\end{align*}
where the last equality holds by noticing that $\frac{\bar{\beta}}{T\left(M_{j}\right)}\rightarrow_{p} 1$ from the law of large numbers and $\left\|\hat{m}-\bar{m}\right\|_{\infty}={\scriptstyle{\mathcal{O}}}_{a.s.}\left(n^{-1 / 2}\right) $ from (\ref{THM:splinemean}).

The estimation error of spatial covariance can be computed as
\begin{align*}
&\|\hat{G}-\bar{G}\|_{\infty}=\max_{1\leq j, j^{\prime}\leq d} \left\|\hat{G}_{jj^{\prime}}-\bar{G}_{jj^{\prime}}\right\|\\
=&\frac{1}{n}\max_{1\leq j, j^{\prime}\leq d}\left\| \frac{\bar{\beta}^{2}}{T\left(M_{j}\right)T\left(M_{j^{\prime}}\right)}\sum_{i=1}^{n}\left(h_{ij}-\bar{h}_{j}\right)\left(h_{ij^{\prime}}-\bar{h}_{j^{\prime}}\right)-\sum_{i=1}^{n}\left(x_{ij}-\bar{m}_{j}\right)\left(x_{ij^{\prime}}-\bar{m}_{j^{\prime}}\right)\right\|\\
\leq &\frac{1}{n}\max_{1\leq j, j^{\prime}\leq d}\left\| \frac{\bar{\beta}^{2}}{T\left(M_{j}\right)T\left(M_{j^{\prime}}\right)}\sum_{i=1}^{n}\left(h_{ij}-\bar{h}_{j}\right)\left(h_{ij^{\prime}}-\bar{h}_{j^{\prime}}\right)-\frac{\bar{\beta}_{j}\bar{\beta}_{j^{\prime}}}{T\left(M_{j}\right)T\left(M_{j^{\prime}}\right)}\sum_{i=1}^{n}\left(x_{ij}-\bar{m}_{j}\right)\left(x_{ij^{\prime}}-\bar{m}_{j^{\prime}}\right)\right\|\\
&+\frac{1}{n}\max_{1\leq j, j^{\prime}\leq d}\left\|\frac{\bar{\beta}^{2}}{T\left(M_{j}\right)T\left(M_{j^{\prime}}\right)}\sum_{i=1}^{n}\left(x_{ij}-\bar{m}_{j}\right)\left(x_{ij^{\prime}}-\bar{m}_{j^{\prime}}\right)-\sum_{i=1}^{n}\left(x_{ij}-\bar{m}_{j}\right)\left(x_{ij^{\prime}}-\bar{m}_{j^{\prime}}\right)\right\|\\
\leq &\max_{1\leq j, j^{\prime}\leq d}\frac{\bar{\beta}^{2}}{T\left(M_{j}\right)T\left(M_{j^{\prime}}\right)}\left\|\hat{G}_{jj^{\prime}}-\bar{G}_{jj^{\prime}}\right\|+\max_{1\leq j, j^{\prime}\leq d}\left(\frac{\bar{\beta}^{2}}{T\left(M_{j}\right)T\left(M_{j^{\prime}}\right)}-1\right)\left\|\bar{G}_{jj^{\prime}}\right\|\\
\leq &\mathcal{O}_{p}\left(1\right)\left\|\hat{G}-\bar{G}|\right\|_{\infty}+{\scriptstyle{\mathcal{O}}}_{p}\left(n^{-1/2}\right)\left\|\bar{G}\right\|_{\infty}\\
=&\mathcal{O}_{p}\left(n^{-1/2}\right)
\end{align*}
where the last equality holds by noticing that $\frac{\bar{\beta}^{2}}{T\left(M_{j}\right)T\left(M_{j^{\prime}}\right)}\rightarrow_{p} 1$ from the law of large numbers and $\left\|\hat{G}-\bar{G}\right\|_{\infty}=\mathcal{O}_{p}\left(n^{-1 / 2}\right) $ from (\ref{THM:splinecov}).

\subsection{Proof of Theorem \ref{THM:FPC}}
\textsc{Proof.}
Denote $\Delta \psi_{k}(\boldsymbol{z})=\int(\hat{G}-G)\left(t, t^{\prime}\right) \psi_{k}\left(t^{\prime}\right) d t^{\prime}$. We have obtained that $\|\hat{G}-G\|_{\infty}=\mathcal{O}_{p}\left(n^{-1/2}\right)$. Thus, for any $k \geq 1,\left\|\Delta \psi_{k}\right\|_{\infty}=\mathcal{O}_{p}\left(n^{-1/2}\right)$. Let
\begin{align*}
\|\Delta\|_{2}=\left\{\iint\left(\hat{G}\left(t, t^{\prime}\right)-G\left(t, t^{\prime}\right)\right)^{2} dt dt^{\prime}\right\}^{1 / 2}=\mathcal{O}_{p}\left(n^{-1/2}\right),
\end{align*}
then according to \cite{Hall06},
\begin{align*}
\hat{\psi}_{k}-\psi_{k}=\sum_{j:j \neq k}\left(\lambda_{k}-\lambda_{j}\right)^{-1}\left\langle\Delta \psi_{k}, \psi_{j}\right\rangle \psi_{j}+\mathcal{O}\left(\|\Delta\|_{2}^{2}\right).
\end{align*}
It follows from Bessel's inequality that
\begin{align*}
\left\|\hat{\psi}_{k}-\psi_{k}\right\|_{2} \leq C\left(\left\|\Delta \psi_{k}\right\|_{\infty}+\mathcal{O}\left(\|\Delta\|_{2}^{2}\right)\right)=\mathcal{O}_{p}\left(n^{-1/2}\right).
\end{align*}
By (2.9) in \cite{Hall06} and $\|\hat{G}-G\|_{\infty}=\mathcal{O}_{p}\left(n^{-1/2}\right)$,
\begin{align*}
\hat{\lambda}_{k}-\lambda_{k}=\iint\left(\hat{G}-G\right)\left(t, t^{\prime}\right) \psi_{k}\left(t\right) \psi_{k}\left(t^{\prime}\right) d t d t^{\prime}+\mathcal{O}\left(\left\|\Delta \psi_{k}\right\|_{2}^{2}\right)=\mathcal{O}_{p}\left(n^{-1/2}\right).
\end{align*}
Next, note that
\begin{align*}
\hat{\lambda}_{k} \hat{\psi}_{k}(t)-\lambda_{k} \psi_{k}(t)=&\int \hat{G}\left(t, t^{\prime}\right) \hat{\psi}_{k}\left(x^{\prime}\right) d t^{\prime}-\int G\left(t, t^{\prime}\right) \psi_{k}\left(t^{\prime}\right) d t^{\prime} \\
=&\int\left(\hat{G}-G\right)\left(t, t^{\prime}\right)\left(\hat{\psi}\left(t^{\prime}\right)-\psi_{k}\left(t^{\prime}\right)\right) d t^{\prime}+\int(\hat{G}-G)\left(t, t^{\prime}\right) \psi_{k}\left(t^{\prime}\right) d t^{\prime} \\
&+\int G\left(t, t^{\prime}\right)\left\{\hat{\psi}_{k}\left(t^{\prime}\right)-\psi_{k}\left(t^{\prime}\right)\right\} d t^{\prime}
\end{align*}
By Cauchy-Schwarz inequality,
\begin{align*}
&\int G\left(t, t^{\prime}\right)\left\{\hat{\psi}_{k}\left(t^{\prime}\right)-\psi_{k}\left(t^{\prime}\right)\right\} d t^{\prime} \leq\left(\int G^{2}\left(t, t^{\prime}\right) d t^{\prime}\right)^{1 / 2}\left\|\hat{\psi}_{k}-\psi_{k}\right\|_{2}=\mathcal{O}_{p}\left(n^{-1/2}\right) \\
&\int(\hat{G}-G)\left(t, t^{\prime}\right)\left(\hat{\psi}\left(t^{\prime}\right)-\psi_{k}\left(t^{\prime}\right)\right) d t^{\prime} \leq\|\hat{G}-G\|_{\infty}\left\|\hat{\psi}_{k}-\psi_{k}\right\|_{2}=\mathcal{O}_{p}\left(n^{-1}\right)\\
&\int(\hat{G}-G)\left(t, t^{\prime}\right) \psi_{k}\left(x^{\prime}\right) d t^{\prime} \leq\|\hat{G}-G\|_{\infty}\left\|\psi_{k}\right\|_{2}=\mathcal{O}_{p}\left(n^{-1/2}\right).
\end{align*}
Therefore, $\left\|\hat{\lambda}_{k} \hat{\psi}_{k}-\lambda_{k} \psi_{k}\right\|_{\infty}=\mathcal{O}_{p}\left(n^{-1/2}\right)$. Then according to $\lambda_{k}\left(\hat{\psi}_{k}-\psi_{k}\right)=\left(\lambda_{k}\hat{\psi}_{k}-\hat{\lambda}_{k}\hat{\psi}_{k}\right)+\left(\hat{\lambda}_{k}\hat{\psi}_{k}-\lambda_{k}\psi_{k}\right)$, we obtain that
\begin{align*}
\lambda_{k}\left\|\hat{\psi}_{k}-\psi_{k}\right\|_{\infty} \leq\left\|\hat{\lambda}_{k} \hat{\psi}_{k}-\lambda_{k} \psi_{k}\right\|_{\infty}+\left|\hat{\lambda}_{k}-\lambda_{k}\right|\left\|\hat{\psi}_{k}\right\|_{\infty}=\mathcal{O}_{p}\left(n^{-1/2}\right).
\end{align*}
It follows that $\left\|\hat{\psi}_{k}-\psi_{k}\right\|_{\infty}=\mathcal{O}_{p}\left(n^{-1/2}\right)$.

\subsection{Proof of Corollary \ref{THM:FPCscore}}
\textsc{Proof.}
For $1\leq i\leq n$, $\hat{\xi}_{ik}-\xi _{ik}$ can be divided into two parts:
\begin{align*}
&R_{1}=\hat{\lambda}_{k}^{-1/2}\int_{0}^{1} \left\{h_{i}
\left(t\right)-\hat{m}\left(t\right) \right\} \hat{\psi}_{k}\left(t\right) dt-\hat{\lambda}_{k}^{-1/2}\int_{0}^{1} \left\{x_{i}
\left(t\right)-m\left(t\right) \right\} \psi_{k}\left(t\right) dt\\
&R_{2}=\hat{\lambda}_{k}^{-1/2}\int_{0}^{1} \left\{x_{i}
\left(t\right)-m\left(t\right) \right\} \psi_{k}\left(t\right) dx-\lambda_{k}^{-1/2}\int_{0}^{1} \left\{x_{i}
\left(t\right)-m\left(t\right) \right\} \psi_{k}\left(t\right) dt
\end{align*}
One assumes that for $k\in\mathbb{N}$, $\lambda_{k}>0$, $\hat{\lambda}_{k}>0$, $\left\|x_{i}-m\right\|_{\infty}$ and the fact that $\left\|\hat{m}-m\right\|_{\infty}=\mathcal{O}_{p}\left(n^{-1/2}\right)$. Moreover, Lemma \ref{xhat} implies that $\left\|h_{i}-x_{i}\right\|_{\infty}={\scriptstyle{\mathcal{O}}}_{a.s.}\left(n^{-1/2}\right)$.
Hence, combining with $\left(\ref{addthree}\right)$, one obtains
\begin{align*}
R_{1}=&\hat{\lambda}_{k}^{-1/2}\int_{0}^{1} \left\{h_{i}
\left(t\right)-\hat{m}\left(t\right) \right\} \left\{\hat{\psi}_{k}\left(t\right)-\psi_{k}\left(t\right)\right\}dt\\
&+\hat{\lambda}_{k}^{-1/2}\int_{0}^{1}\left\{h_{i}
\left(t\right)-\hat{m}\left(t\right)-x_{i}\left(t\right)+m\left(t\right)\right\}\psi_{k}\left(t\right)dt\\
\leq&\hat{\lambda}_{k}^{-1/2}\left\|x_{i}-m\right\|_{\infty}\left\|\hat{\psi}_{k}-\psi_{k}\right\|_{\infty}+\hat{\lambda}_{k}^{-1/2}\left(\left\|h_{i}-x_{i}\right\|_{\infty}+\left\|\hat{m}-m\right\|_{\infty}\right)\left\|\psi_{k}\right\|_{\infty}\\
=&\mathcal{O}_{p}\left(n^{-1/2}\right).
\end{align*}
Through first order Taylor expansion of $\hat{\lambda}_{k}$ at $\lambda_{k}$, one gets $\hat{\lambda}_{k}^{-1/2}=\lambda_{k}^{-1/2}-\left(1/2\right)\lambda_{k}^{-3/2}\left(\hat{\lambda}_{k}-\lambda_{k}\right)+{\scriptstyle{\mathcal{O}}}\left(\left|\hat{\lambda}_{k}-\lambda_{k}\right|\right)$. Hence, $\left(\ref{addone}\right)$ ensures that $\left|\hat{\lambda}^{-1/2}_{k}-\lambda_{k}^{-1/2}\right|=\mathcal{O}_{p}\left(n^{-1/2}\right)$. Consequently,
\begin{align*}
R_{2}=&\left(\hat{\lambda}^{-1/2}_{k}-\lambda_{k}^{-1/2}\right)\int_{0}^{1} \left\{x_{i}
\left(t\right)-m\left(t\right) \right\} \psi_{k}\left(t\right) dt\\
\leq&\left|\hat{\lambda}^{-1/2}_{k}-\lambda_{k}^{-1/2}\right|\left\|x_{i}-m\right\|_{\infty}\left\|\psi_{k}\right\|_{\infty}=\mathcal{O}_{p}\left(n^{-1/2}\right).
\end{align*}
Then theorem \ref{THM:FPCscore} is proved by $\max_{1\leq i\leq n}\left\|\hat{\xi}_{ik}-\xi_{ik}\right\|=\max_{1\leq i\leq n}\left(\left\|R_{1}\right\|+\left\|R_{2}\right\|\right)=\mathcal{O}_{p}\left(n^{-1/2}\right)$.

\end{document}